\documentclass[12pt]{article}
\usepackage[dvips]{epsfig}
\usepackage{latexsym}
\usepackage[bf,footnotesize]{caption}	
\begin{document}
\setlength{\unitlength}{1mm}


\newcommand{\ba} {\begin{eqnarray}}
\newcommand{\ea} {\end{eqnarray}}
\newcommand{\be}{\begin{equation}}
\newcommand{\ee}{\end{equation}}
\newcommand{\n}[1]{\label{#1}}
\newcommand{\eq}[1]{Eq.(\ref{#1})}
\newcommand{\ind}[1]{\mbox{\tiny{#1}}}
\renewcommand\theequation{\thesection.\arabic{equation}}

\newcommand{\nn}{\nonumber \\ \nonumber \\}
\newcommand{\nl}{\\  \nonumber \\}
\newcommand{\pr}{\partial}
\renewcommand{\vec}[1]{\mbox{\boldmath$#1$}}

\newcommand{\ben}{$$}
\newcommand{\een}{$$}
\newcommand{\bea}{\begin{eqnarray}}
\newcommand{\eea}{\end{eqnarray}}
\newcommand{\bean}{\begin{eqnarray*}}
\newcommand{\eean}{\end{eqnarray*}}
\newcommand{\e}{{\rm e}}
\newcommand{\tr}{{\rm tr}}
\newcommand{\av}[1]{<\!{#1}\!>_{\omega}}

\title{ Stochastically Fluctuating Black-Hole Geometry, Hawking
Radiation and the Trans-Planckian Problem}
\author{\\
C. Barrab\`{e}s\thanks{e-mail: barrabes@celfi.phys.univ-tours.fr}  ${}^{1}$,
V. Frolov\thanks{e-mail: frolov@phys.ualberta.ca}
${}^{2, 1}$ and 
R. Parentani\thanks{e-mail: 
parenta@celfi.phys.univ-tours.fr}
${}^{1}$ 
\date{\today}}
\maketitle
\noindent 
{
\\ $^{1}${\em Laboratoire de Math\'{e}matiques et Physique
Th\'{e}orique, 
CNRS UPRES A 6083,\\${ }\quad$
Universit\'{e} de Tours, 
37200 Tours, France}\\
$^{2}${ \em
Theoretical Physics Institute, Department of Physics, \ University of
Alberta, \\ ${ }\quad$ Edmonton, Canada T6G 2J1}
}
\bigskip

\begin{abstract}
We study the propagation of null rays and massless fields in a black
hole  fluctuating geometry. The metric fluctuations are induced by a
small oscillating incoming flux of energy. The flux also induces
black hole mass oscillations around its average value. We assume that the
metric fluctuations are described by a statistical ensemble. The stochastic
variables are the phases and the amplitudes of Fourier modes of the
fluctuations.  By averaging over these variables, we obtain an
effective propagation for massless fields which  is characterized by a
critical length  defined by the amplitude of the metric fluctuations:
Smooth wave packets with respect to this length are not  significantly
affected when they are propagated forward in time.  Concomitantly, we
find that the asymptotic  properties of Hawking radiation are not
severely modified. However, backward propagated wave packets are
dissipated by the metric fluctuations once their blue  shifted
frequency reaches the inverse critical length. All these  properties
bear many resemblences with those obtained in models for black hole
radiation based on a modified dispersion relation.  This strongly
suggests that the physical origin of these models,  which were
introduced to confront the trans-Planckian problem, comes from 
the fluctuations of the black hole geometry.

\end{abstract}
\vspace{.3cm}


\newpage

\section{Introduction}

In his original derivation of black hole radiance,  Hawking
\cite{Hawk:75} considered the propagation of a linear quantized field
in a  classical background geometry, that of a collapsing body. In
this framework, one neglects two effects. One first neglects the
gravitational back reaction effects, i.e. the consequences of the
quantum response of the geometry to the energy density of the
radiation field.  To compute these effects is at present out of reach
as it requires a better understanding of quantum gravity. 

One also neglects the fluctuations of the geometry which are not due to
the energy density of the radiation field. Besides quantum
mechanical fluctuations of the gravitational field itself, there exist
also metric fluctuations induced by the quantum fluctuations
of other fields. The latter can be approximatively described 
by introducing stochastic noice sources 
in the right-hand side of the Einstein equations
\cite{Hu:99,CaHu:98,MaVe:98a,MaVe:98b}. 
In this description, one therefore deals
with a stochastic ensemble of fluctuating geometries. 
Our aim is to study the propagation of a massless
field in such an ensemble.  

To describe metric fluctuations near the black hole horizon  we shall
use a model similar to that considered by York
\cite{York:83}. It is based on the hypothesis that the metric 
fluctuations are driven by a small oscillating flux of energy 
of an infalling null fluid. In our model, the metric fluctuations are 
represented by a linear superposition  with different  frequencies. 
Stochasticity  will come into the picture by
assuming that the amplitudes and phases of each mode are
stochastic variables. Therefore to obtain the expectation value of any
observable will require averaging over these variables. As we shall
still neglect gravitational back reaction effects, we shall still have
a linear field equation. In a former paper \cite{BaFrPa:99} , we  analyzed the
propagation of a test field in a metric characterized by a given
classical fluctuation. Therefore, the novelty of the present work is
to take an ensemble of such fluctuations. 

The influence of black hole metric fluctuations on physical effects in
the black hole geometry is an interesting and  open problem.
As we shall demonstrate in the paper, one of the justifications of 
considering a statistical  ensemble of metric
fluctuations is to confront the trans-Planckian  problem\cite{Unruh81,
Jac91, Jac93, tHooft, VV, MP95, thooft96, Unruh95, BMPS95, Jac96}.  
This problem
comes from the fact that low energy modes reaching ${\cal{J}}^+$ emerge
from field configurations which possess, near the event horizon,
arbitrarily high frequencies (as measured in a freely falling frame).
Since gravitational  interactions grow with the energy, one must call
into  question the validity of describing the propagation of these
field configurations by free field theory. In particular, one might 
wonder if the low energy asymptotic properties  of Hawking radiation,
namely stationarity and thermality with a temperature determined by the
surface gravity, are sensitive to the high energy behavior of the
theory.

To answer this point, Unruh made an interesting proposal 
\cite{Unruh81,Unruh95}. 
He considered the propagation of sonic waves in an acoustic geometry
governed by a wave equation whose dispersion relation is bended at high
frequency. Indeed, only low frequency phonons
propagate freely with a given velocity. For frequencies higher than
a critical frequency $\omega_c$,  the dispersion relation is no
longer linear and dissipation may occur. He then showed that the
modification of the dispersion relation in no way affects the
asymptotic properties  of Hawking radiation as long as  $\omega_c$ is
much bigger than the surface gravity $\kappa$.  The reason is that
there is an adiabatic decoupling between  these two energy
scales. In \cite{BMPS95}, it was conjectured that light propagation 
 near a black hole horizon should also be described by 
an effective mutilated theory and an alternative model 
``which may be more appropriate to black hole physics''
was proposed.
The basic argument is that when a Hawking quantum 
is propagated backward close to the
horizon, it will interact with a reservoir of modes, e.g. the high
angular momentum modes which do not reach spatial infinity, or the
metric fluctuations induced by these modes \cite{CEMNP}.

In this paper, we consider a stochastic ensemble\footnote{
There are several works in the literature wherein 
stochastic ensembles of black-hole metrics have been considered, 
see e.g. \cite{HuSh:98,WuFo:99}.
 However, to our knowledge, none of them confronts
the role of ultra-high frequencies occuring in Hawking radiation.}
of metric fluctuations to describe these interactions.
We shall show that light propagation in a stochastic metric indeed
leads to an effective truncated theory near the event horizon. More
precisely, we obtain the following.  First the critical lenght
$\omega_c^{-1}$ is determined by the amplitude of the metric
fluctuations. Secondly, as far as forward propagation is concerned, 
the evolution of smooth wave packets (where smooth means that their
in-frequency content is much below $\omega_c$) is affected only slightly
by the metric fluctuations.  Thirdly, backward in time propagation of
wave packets representing Hawking quanta is dramatically modified
only when the blue shift factor brings their frequency close to
$\omega_c$. {In this regime, the amplitude of the wave packet is
rapidely dissipated (backward in time!).}

At this point, the reader might wonder about the physical validity
of  these results since we do not know the precise nature of the
metric fluctuations near the event horizon. In this respect, the
following point should be emphasized. We are not studying the
fluctuations themselves but only their effects on light propagation. 
As argued by Feynman and Hibbs\cite{FH}, the
effective propagation obtained by tracing out the degrees of freedom
of the environment  does not depend on the  precise nature of the
interactions of the test particle with it. For this universality to
apply, one should neither ask too detailed questions (e.g. during too
small time intervals) nor consider too strong interactions leading to
significant recoils effects in the
environment. The characterization of the domain of validity
of the stochastic treatment is a complicated question which
requires a detailed knowledge of the environment 
dynamics\cite{VanK}. 
In the case of metric fluctuations, this is of course
beyond our reach. But the crucial point is that, if there is an
intermediate  regime in which  the particle weakly interacts with a
large number of modes, it is {\it sufficient} to work with
an appropriate simplified model.
In our case, universality comes essentially from
the  exponentially growing Doppler effect encountered in  backward
propagation: how ever small is the critical length it will be reached
in a logarithmicly short (advanced) time. The only condition it must
obey is to be much  smaller than the Schwarzschild radius.

The paper is organized as follow. In Section 2, we describe our
fluctuating metric and we study how outgoing null rays propagate in it. 
The metric we choose results from the collapse  of a massive
spherically symmetric null shell which is followed by a small
additional oscillating  and infalling null flux.

In Section 3, we analyse the propagation of wave packets of  a
massless scalar field in an ensemble of fluctuating metrics. That is,  we
first obtain the propagation of a given initial wave packet in each
realization of the geometry and then define the mean wave packet
by averaging over the fluctuating  part of the metric.  Since the
linearity of the problem is maintained,  we formulate the problem as
in a S-matrix language and show that this matrix is  diagonal
in Fourier components. From
this equation one immediately sees that only the high frequency part
of the spectrum is affected by the fluctuations. 

In Section 4, we first analyse the Green function in the  initial
vacuum and show that the metric fluctuations do not significantly
affect its short distance behavior thereby guaranteeing unmodified
properties of Hawking radiation. This is verified by computing
directly the asymptotic flux of energy of quantum radiation. 

In Section 5, we compare the backward propagations from ${\cal{J}}^+$
of a wave function  describing a typical Hawking quantum in two
cases: in our fluctuating metric and by modifying the dispersion
relation as done in \cite{Unruh95, BMPS95}. The similarities are
manisfest. We conclude the paper by making comments on
the apparent breakdown  of Lorentz invariance which
appears in these effective theories.


\section{Fluctuating Black-Hole Geometry}
\setcounter{equation}0

\subsection{Metric ansatz}
In this article we shall only consider spherical  modes of
metric fluctuations propagating in a spherically symmetric background. 
The most general spherical metric can be written in the
form 
\be
dS^2=\gamma_{AB}\,dx^A\,dx^B+R^2\,d\omega_2^2\,, \n{2.1}
\ee
where $A,B=0,1$, and $\gamma_{AB}$ and $R$ are functions of $x^A$. 
The metric of a black hole of mass $M$ formed by the
collapse at $v=0$ of a massive null shell with mass $M$ is   
\be
dS^2=(4M)^2 ds^2\, ,\hspace{0.5cm}
ds^2 = -A\,dv^2+2dv\,dr+r^2d\omega_2^2\,, \n{2.2}
\ee
where in the absence of fluctuations
\be
A=A_0(r,v) = 1-\frac{\vartheta (v)}{2r}\, , \n{2.3}
\ee 
$\vartheta (v)$ being the Heaviside step function equal to 1 for
positive argument. For further convenience we have introduced the
dimensionless coordinates $(v,r)$, so that $R=4Mr$ and $4Mv$ are the
radius and the advanced time in units where $G=c=1$. A conformal
Penrose-Carter diagram of the spacetime is shown in Figure~1.

\begin{figure}
\centerline{\epsfig{file=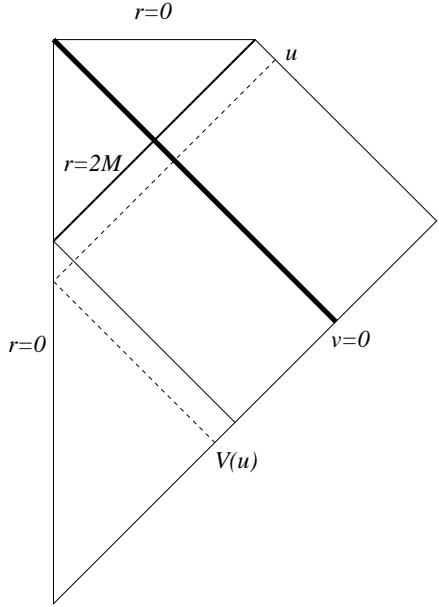, height=8cm}}
\caption[st]{Conformal Penrose-Carter diagram of the unperturbed 
collapsing black hole geometry. The massive (of mass $M$) null shell is
represented by a heavy solid line at $v=0$.}
\label{st}
\end{figure}

The most general metric perturbation preserving the form (\ref{2.1}) of the
metric is described by four functions of $x^A$:
$\delta r$ and $\delta\gamma_{AB}$. The remaining coordinate gauge
freedom is generated by infinitesimal coordinate transformations
$\xi^A(x)$. We fix the gauge by putting
\be
\delta r=0\,, \qquad \delta\gamma_{rr}=0\,. \n{2.4}
\ee
The perturbed metric can be written in the form
\be
ds^2 = \Psi (-A\,dv^2+2dv\,dr)+r^2d\omega_2^2\,, \n{2.5}
\ee
with
\be
\Psi = 1+\delta\Psi \,, \qquad A=A_0+\delta A\,. \n{2.6}
\ee 

Upon restricting attention to the propagation of radial null rays,
the 2-dimensional conformal factor $\Psi$ plays no role.  In the
following sections, we shall study the propagation of s-waves 
in the fluctuating black hole geometry. We shall demonstrate
that, for $s$-modes, $\Psi$ also drops out of the 4-dimensional
Dalembertian. Hence only the function $A$ will be relevant for us.

To further simplify the problem,  
we assume that the metric fluctuation $\delta A$ is composed
only of infalling radial null modes. Thus it is of the form
\be \n{2.7}
\delta A = - \frac{1}{2r}\,\vartheta (v)\, \mu (v)\, .
\ee
so that the perturbed metric is given by (\ref{2.5}) with
\be
\n{2.8}
A=A_0 + \delta A =1-{\vartheta (v) [1+ \mu (v)] \over 2 r}\, .
\ee
For $\Psi=1$ this is a Vaidya metric.  The function $\mu (v)$ encodes
the light-like infalling fluctuations.  As in eq. (\ref{2.3}), the step
function in relation (\ref{2.8})  indicates that the black hole
results from the gravitational collapse at  $v=0$ of a massive  null
shell with mass $M$, and that there are no fluctuations prior  to the
collapse of the null shell. 
Therefore spacetime is flat to the past of the null shell.

\subsection{Stochastic variables}

To introduce the stochastic variables in simple terms,
we postulate that $\mu (v)$
possesses a discrete\footnote{
A discrete spectrum arises for example in  York's approach \cite{York:83}
based on the quasi-normal modes of the black-hole metric. 
However, the spectrum due to other fields can be continuous. 
The results of the present paper can be easily adopted to this case. 
It is sufficient to replace the discrete sum, $\sum_\omega$, 
by an integral, $\int\, d\omega\, \nu(\omega)$, 
where $\nu(\omega)$ is the number density of
fluctuation modes.} 
and non-degenerate Fourier decomposition:
\be \n{2.9}
\mu(v)=\sum_{\omega} \left[ \mu_1^{\omega} \sin(\omega v)+
\mu_2^{\omega} \cos(\omega v)\right]
=
\sum_{\omega} \mu_0^{\omega}\,\sin (\omega
v+\phi_{\omega})\, . 
\ee  
The equality is obtained by using polar
coordinates in the $(\mu_1,\mu_2)-$plane
\be \n{2.10}
\mu_0^{\omega} =\sqrt{(\mu_1^{\omega})^2+(\mu_2^{\omega})^2}\, ,\hspace{0.5cm}
\tan\phi_{\omega} = {\mu_2^{\omega} \over \mu_1^{\omega}} \, .
\ee

We assume that the (real) amplitudes 
$\mu_1^{\omega}$ and $\mu_2^{\omega}$
are stochastic variables taking range from $-\infty$ to $\infty$.
For simplicity we postulate that they are stochastically 
independent and that, for any frequency $\omega$, the
dispersions of $\mu_1^{\omega}$ and $\mu_2^{\omega}$ are equal.
Thus, there is no preferred value of
the phase $\phi_{\omega}$ in the $(\mu_1^{\omega},\mu_2^{\omega})$ plane.
In this case, $\tilde{\rho}_{\omega}(\mu_0^{\omega})$, the distribution 
function for the amplitude $\mu_0^{\omega}$, satisfies the normalization
condition 
\be \n{2.11}
\int_{0}^{2\pi}\, d\phi_{\omega} \,\,\int_{0}^{\infty}\,d\mu_0^{\omega}\,
\mu_0^{\omega}\,\tilde{\rho}_{\omega}(\mu_0^{\omega})\,=\,1\,.
\ee 
Later in the text, we shall assume that the amplitude $\mu_0^{\omega}$ 
is a Gaussian variable whose distribution function 
is equal to  
\be \n{2.12}
\tilde{\rho}_{\omega}(\mu_0^{\omega})\,=\, 
{1\over 2 \pi \tilde{\sigma}_{\omega}^2}\,
\exp \left[-{(\mu_0^{\omega})^2 \over 2 \tilde{\sigma}_{\omega}^2} \right]\,.
\ee
The coefficient $\tilde{\sigma}_{\omega}$ determines the dispersion of the
amplitude $\mu_0^{\omega}$:
\be \n{2.13}
<(\mu_0^{\omega})^2>_{\omega}\,=\, 2\tilde{\sigma}^2_{\omega}\, ,
\ee
where $< >_{\omega}$ represents the average calculated with
the distribution  $\tilde{\rho}_{\omega}(\mu_0^{\omega})$.

In what follows we shall have to deal with observables 
depending on the fluctuating geometry which obey the following
factorization condition
\be \n{2.14}
Q=\prod_{\omega} Q_{\omega}(\mu_0^{\omega},\phi_{\omega})\, .
\ee
For these observables, we can consider each sector 
labeled by $\omega$ separately. It is then usefull to 
introduce the successive averages:
\be \n{2.15}
\bar{Q}_{\omega}(\mu_0^{\omega})={1\over 2\pi}\int_0^{2\pi}\,\,d\phi_{\omega}
\,\,Q_{\omega}(\mu_0^{\omega},\phi_{\omega})\, ,
\ee
\be \n{2.16}
\av{Q_{\omega}}=2\pi \int_{0}^{\infty}\,\, d\mu_0^{\omega}\,\, \mu_0^{\omega}\,\, 
\tilde \rho_{\omega}(\mu_0^{\omega}) \,\,\bar{Q}_{\omega}(\mu_0^{\omega})\, ,
\ee
\be \n{2.17}
\ll\! Q\!\gg=\prod_\omega \av{Q_{\omega}}\, .
\ee
The first equality gives $\bar{Q}_{\omega}$, the average of  $Q_\omega$
over the stochastic phase $\phi_{\omega}$. The second one  gives the
result of averaging   $\bar{Q}_{\omega}$ over the amplitude
$\mu_0^{\omega}$. Finally, the overall ensemble average of the
observable $Q$,  is given by (\ref{2.17}). The order in this averaging
procedure follows from the fact that to perform the first average, one
simply has to assume that there is no prefered direction in
$\phi_\omega$. For the second instead, we need to know the distribution
$\tilde \rho_\omega$. And for the third one, we must know the whole
spectrum.

To determine the actual physical spectrum of metric fluctuations 
around a black hole horizon is an extremely complicated problem
requiring  a theory of quantum gravity, see \cite{York:83, CEMNP} for
attemps to characterize this spectrum. In the present paper  we shall
not use any specific form of the spectrum. As we already mentioned, 
the effects of the metric fluctuations
hardly depend on its exact form. That this is the case will appear 
progressively in the paper. Our only assumption is that the
amplitute of metric fluctuations are much smaller than the gravitational
radius of the black hole. In our dimensionless units, this gives
$\tilde \sigma \ll 1$. This should be true for black
holes of mass $M$ much greater than the Planckin mass $m_{\ind{Pl}}$
(in \cite{York:83}, the estimate dimensionless amplitude 
scales as $\tilde{\sigma}\sim (m_{\ind{Pl}}/M)$, 
whereas in \cite{CEMNP} $\tilde{\sigma}\sim (m_{\ind{Pl}}/M)^{4/3}$).

Because of the factorizability of the operators  and the
statistical independence
of the amplitudes,  it will be sufficient to consider only a single
fluctuation mode. To simplify the notations we shall drop the index
$\omega$  in the amplitude and in the phase. That is  we shall work  in
the fluctuating geometry 
(\ref{2.2}), (\ref{2.8}) with
\be 
\mu(v)= \mu_0\,\sin (\omega v+\phi)\, , \n{2.18}
\ee
with $\mu_0 \ll 1$.  We call the metric
(\ref{2.2}), (\ref{2.8}) with $\mu(v)$ given  by (\ref{2.18}) a 
realization of the fluctuating geometry. By averaging over $\phi$ and
$\mu_0$ we thus assume that we are dealing with an ensemble of such
realizations.

Finally, it should be stressed that $\phi$, the phase of the
fluctuations, has physical meaning. Indeed, the gravitational collapse
of the null shell singles out a special moment of time  (e.g. the
moment of formation of the horizon), and fluctuations with different
phases are not equivalent.  Therefore $\phi$ is a stochastic variable
and  when computing the statistical average, integration over it is to
be done.

\subsection{Null ray propagation in a fluctuating geometry}
To characterize the propagation of radial null rays in a given realization
of the geometry we focus on the relation
\be\n{2.19}
v=V_\phi(u)\ 
\ee
between the moment $v$ of advanced time when the null ray was emitted
from ${\cal J}^-$ and  the retarded time $u$ when it reaches ${\cal
J}^+$.  
The above equation contains a simplified notation since $V_\phi(u)$ 
also depends on the amplitude $\mu_0$ and 
the frequency $\omega$ -- this is
evident in eq. (\ref{2.20}) below.

In the late time regime, i.e. ${u}\gg 1$, 
we have (for details, see 
\cite{BaFrPa:99})
\[
V_\phi(u) = -\left[1+\,\frac{\mu_0}{\sqrt{1+\omega^2}}\sin(\phi
+\phi_0)+\right.
\]
\be \n{2.20}
\left.
e^{-u} \left( 1
+ \frac{\mu_0}{\omega} \cos(\phi+ 2 \phi_0)
+
\,\frac{\mu_0\, q(\omega)}{\sqrt{1+\omega^2}}
\sin(\phi+ \phi_0 + \omega u - \varphi_{\Gamma}(\omega )
)
\right)\right] + O(\mu{_0}{^2})
\, ,
\ee
where $\phi_0 =\arctan \omega$ and where $\varphi_{\Gamma}(\omega )$
and $q(\omega)$ only depend on the frequency of fluctuations.    In
this Section and the next one, we shall work with a simplified version of 
$V_\phi(u)$: Since the last two terms inside the large parenthesis in
(\ref{2.20}) are much smaller than 1 ($\mu_0 \ll 1$), we  shall simply
omit them. In Section 4 instead, we shall take them into account  
and also include all quadratic terms in $\mu_0$.

The simplified version of relations  (\ref{2.19}) and
(\ref{2.20}) is of the form
\be
w=W_\phi(u),\, \n{2.21}
\ee
\be
W_\phi(u)= \,w_0\sin(\phi
+\phi_0)+e^{-u} 
\,,
\n{2.22}
\ee
where we have introduced for later convenience
\be
w = -1 -v\, , \hspace{0.5cm}
w_0=\frac{\mu_0}{\sqrt{1+\omega^2}}\,. \n{2.23}
\ee

For a given realization of the geometry and to first order in
$\mu_0$  the event horizon is given by the equation 
(see eq. (3.10) in \cite{BaFrPa:99})
\be
r^{\ind{EH}}_{\phi}(v) = {1\over 2}[1+w_0\,\sin (\omega v+
\phi+\phi_0 )]\,.
\n{2.24}
\ee
It crosses the collapsing null shell, $v=0$ or $w=-1$, at the radius
\be
r^{\ind{EH}}_{\phi}(0) = {1\over 2}[1+w_0\,\sin (\phi+\phi_0 )]\,.
\n{2.25}
\ee
Being traced backward in time the null geodesic giving rise to the
horizon enters the flat spacetime region inside the collapsing shell,
bounces at $r=0$, and finally reaches ${\cal J}^-$ with $w$
(defined by (\ref{2.23})) lying in the domain
\be
w \in (-w_0,w_0) \,.\n{2.26}
\ee 
Therefore, for all $\phi$, radial
null rays emitted from ${\cal J}^-$ at advanced time $w<-w_0$ fall
into the singularity $r=0$. On the
contrary, radial null rays emitted with $w>w_0$
always reach ${\cal J}^+$. The time of their arrival to ${\cal J}^+$
lies in the interval
\be
u \in (u_-,u_+) \,, \n{2.27}
\ee
where $u_{\pm}$ are
\be
u_{\pm} = - \ln (w \mp w_0) \,. \n{2.28} 
\ee 
Finally, the rays emitted in the interval $-w_0<w<w_0$ reach ${\cal
J}^+$ only for some values of $\phi$. Since for $w$ in this interval
there always exists a phase such that the ray propagates along the
horizon, the moment of arrival at ${\cal J}^+$ varies from $u_- = -\ln
(w+w_0)$ to $u_+ = \infty$.

Figure~2 shows the behavior of radial null rays in the fluctuating 
black hole geometry characterized by a single mode of frequency
$\omega$. Ray {\bf a} represents the event horizon, while rays {\bf b}
and {\bf c} are rays which reach ${\cal J}^+$ or remain trapped inside
the horizon, respectively.  When they are traced back in time all these
rays pass through the center of symmetry $r=0$, and go to ${\cal J}^-$
along  lines $v=const$. It should be noted that the conformal
Penrose-Carter diagrams for any given realization of the fluctuating
black hole geometry is similar to the one shown in Figure~1 even
though  the event horizon no longer obeys the equation $r=1/2$, see eq.
(\ref{2.24}).

\begin{figure}
\centerline{\epsfig{file=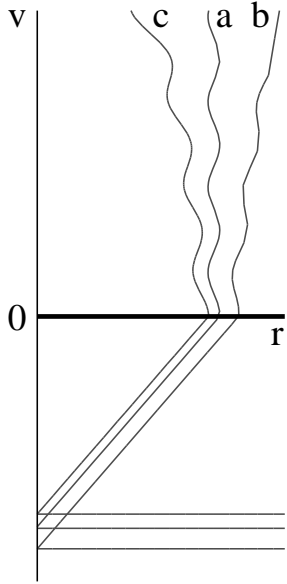, height=8cm}}
\caption[rays]{Null rays propagation in a fluctuating geometry of the
black hole.}
\label{rays}
\end{figure}

\section{Wave Propagation in a Fluctuating Geometry}\label{s2}
\setcounter{equation}0

\subsection{$\delta$-pulse propagation}

\subsubsection{Wave propagation in a given realization of the geometry}
In this article, we study only $s$-modes of a minimally coupled
massless scalar field $\chi$.  We introduce as usual $\varphi = r \chi$.
Then, the four-dimensional  Dalembertian equation $\Box\chi =  0$
when computed in our fluctuating metric (\ref{2.5}) gives, see
e.g \cite{FrNo:98},
\be
\left[{}^{(2)\!}\Box - {\partial_r A \over r }
\right]\varphi =  0 \n{3.1}
\ee
Hence $\Psi$ defined in eq. (\ref{2.5}) plays no role.
Moreover, upon neglecting the centrifugal quantum potential,
one obtains the equation
\be
{}^{(2)\!}\Box\varphi =  2 \partial_v \partial_r \varphi+ 
\partial_r (A \partial_r \varphi)  =  0 \, .\n{3.2}
\ee
When adopting this equation, one works
in the geometrical optics approximation. Therefore, the
spherically symmetric perturbations of the
metric affect the global properties of the solutions of (\ref{3.2}) 
only through the glueing of the null characteristics
encoded in $V_\phi(u)$ given in eq. (\ref{2.22}).
By ``global properties'' we mean properties which
can be measured on ${\cal J}^+$. We shall not compute 
the local value of the field near the event horizon in the
fluctuating geometry. This would require the knowledge 
of the local description of outgoing null geodesics
$u_\phi(v,r)$ for arbitrary $v$ and $r$. Even though we restrict
ourselves to global aspects, we shall see that the 
notion of a smeared horizon naturaly emerges. This 
should cause no surprise since the very definition
of event horizon is global. 

We shall denote the value of the solutions of (\ref{3.2})
on ${\cal J}^{\pm}$ by a capital  letter
\[
\Phi^{\pm}=  \varphi|_{{\cal J}^{\pm}}\, .
\]
Using the coordinate $w$ defined in (\ref{2.23}) we call $\Phi^-(w)$
the initial value (or image) of the solution $\varphi$ on ${\cal
J}^-$, and $\Phi^+(u)$ the final value (or image) of $\varphi$ on
${\cal J}^+$. For a fixed geometry, i.e. for a fixed $\mu_0$ and $\phi$,  the
knowledge of $\Phi^-(w)$ uniquely determines 
its image on ${\cal J}^+$.
However, for backward propagation from ${\cal J}^+$ to ${\cal J}^-$, 
this is not true since ${\cal{J}}^+$ is not a complete Cauchy surface.

Let us consider what happens to a wave packet propagating in  the
black-hole fluctuating geometry.  The wave packet is sent from ${\cal
J}^-$ where it has image $\Phi^-(w)$.  Because of the linearity of
the problem, it is sufficient to know the evolution of the following
$\delta$-like pulse sent from ${\cal J}^-$
\be
\Delta^-(w|w') = \delta (w-w')\,. \n{3.3}
\ee
We call ${\Delta}^{+}_{\phi}(u|w')$ the image of this pulse on ${\cal
J}^+$ in a given realization characterized by the value of $\phi$.
Since we work in the geometrical optics approximation, it is 
equal to
\be
\Delta^{+}_{\phi}(u|w') = \delta (W_\phi(u)-w')\,, \n{3.5}
\ee  
where $W_\phi(u)$ is the value of  $w$ at which a radial null ray has
to be  sent from ${\cal J}^-$ in order to reach ${\cal J}^+$ at
retarded time $u$. 

In terms of  ${\Delta}^+_{\phi}$, the 
image of the wave packet  on ${\cal J}^+$ in this realization is given by
\be
{\Phi}^+_{\phi}(u)=\int_{-\infty}^{\infty}dw\, {\Phi}^-(w)\, 
{\Delta}^+_{\phi}(u|w)\,. \n{3.4}
\ee

\subsubsection{Averaging over the phase}

Using the notation introduced in eq. (\ref{2.15})  
we denote by $\bar{\Delta}^+(u|w)$ the  average of
$\Delta^{+}_{\phi}(u|w)$ over the phase. 
Throughout the paper the ``bar'' over a quantity indicates 
that this average has been taken. 
In terms of  $\bar{\Delta}^+$, the average
image of the wave packet  on ${\cal J}^+$, is given by
\be
\bar{\Phi}^+(u)=\int_{-\infty}^{\infty}dw\, {\Phi}^-(w)\, 
\bar{\Delta}^+(u|w)\,. \n{3.6}
\ee
Thus our primary goal is to calculate $\bar{\Delta}^+(u|w)$.
By definition it is given by
\be
\bar{\Delta}^+(u|w)=\frac{1}{2\pi}\,\int_0^{2\pi}\delta
(W_\phi(u)-w)\,d\phi\,. \n{3.7}
\ee
We remind the reader that this equation follows from the
hypothesis that all values of $\phi$ are equally probable,
see section 2.2.
To calculate this integral, we use the following property of
$\delta$-function:
\be
\int_0^{2\pi}d\phi \,\delta (F(\phi )) =
\sum_i\frac{1}{|F'(\phi_i)\,|}\,, \n{3.8}
\ee
where $\phi_i$ are the roots of the equation $F(\phi )=0$ lying in the
interval $(0,2\pi )$. That is, we simply need to find those solutions
$\phi$ of the equation $W_\phi(u) = w$ which lie in the interval
$(0,2\pi )$.

Using the simplified expression (\ref{2.22}) we obtain\footnote{Notice 
that $\bar{\Delta}^+$ is highly non linear 
in $w_0$. However it results from eq.  (\ref{2.22})
which has been linearized and simplified. Therefore,
one must question the validity of its non-linear dependence
in $w_0$. This question is addressed in Appendix A.
The outcome of the analysis is that one can trust (\ref{3.9})
in the non-linear regime since it correctly gives
the dominant non-linear effects. The fact that 
the higher order terms in
$\mu_0$ of $W_\phi$ are all irrelevant indicates that there is an 
underlying universality of our results.}
\be
\bar{\Delta}^+(u|w) =\frac{\vartheta (w_0^2 - (e^{-u}- w)^2)}
{\pi \sqrt{w_0^2 -(e^{-u}-w)^2}}\, .\n{3.9}
\ee
For a fixed $w$, expression (\ref{3.9})  determines
$\bar{\Delta}^+(u|w)$ for $u\in (u_-,u_+)$, where
\be
u_- = -\ln (w+w_0)\,, \n{3.10}
\ee
\be
u_+ = \left\{\begin{array}{ll}
-\ln (w-w_0)\,, & \mbox{for } \ w>w_0;\\
\infty \,, & \mbox{for } \ -w_0<w<w_0\,. \end{array}\right. \n{3.11}
\ee
For $u \notin (u_-,u_+)$, $\bar{\Delta}^+(u|w)$ vanishes
since no $\phi \in (0,2\pi )$ satisfies $W_\phi(u)=w$. Similarly, for
a fixed value of $u$, $\bar{\Delta}^+(u|w)$ does not vanish only for
$w \in (-w_0+e^{-u},w_0+e^{-u})$.

In a similar way we can consider propagation backward in time. It is
again sufficient to make the calculations for a $\delta$-like pulse.
We call
\be \n{3.12}
{}^+\Delta(u|u')=\delta(u-u')\, 
\ee
the image of the pulse on ${\cal J}^+$. Then, for a given realization
of the geometry it has the following image on ${\cal J}^-$
\be \n{3.13}
{}^-\Delta_\phi (w|u')=\delta(u'-U_\phi(w))\, .
\ee
Using eq. (\ref{2.22}), one has
\be \n{3.14}
U_\phi(w)=-\ln\left[w-w_0 \sin(\phi +\phi_0)\right]\, .
\ee

By averaging $^-\Delta_\phi$ over $\phi$ we get
\be \n{3.15}
{}^-\bar{\Delta}(w|u)= e^{-u} \bar{\Delta}^+(u|w)  \, , 
\ee
where $\bar{\Delta}^+(u|w)$ is given in (\ref{3.9}).  The only
difference is the Jacobian $e^{-u}$ which relates the delta functions
defined on ${\cal J}^-$ in terms of $w$ and on ${\cal J}^+$ in terms of
$u$. Then, the image on  ${\cal J}^-$ which results from the 
backward propagation of a wavepacket which has the image $\Phi^+$ on
${\cal J}^+$  is given by
\be
{}^-\bar{\Phi}(w)=\int_{-\infty}^{\infty}du\, {\Phi}^+(u)\, 
{}^-\bar{\Delta}(w|u)\,. \n{3.aa}
\ee
It should be noted that in order to get this relation we assumed 
that, for any realization of the geometry, no signal was propagating
backward from black hole interior.
This point will be further discussed in section~3.3.

\subsubsection{Averaging over the amplitude}

The average value of an observable over
an ensemble of amplitudes of metric fluctuation $\mu_0$ 
characterized by the distribution
$\tilde{\rho}_{\omega}(\mu_0)$ is given by
eq. (\ref{2.16}).
In order to compute the amplitude average of $\bar \Delta^+$, 
we need to know $\tilde{\rho}_{\omega}$.
In the case of a Gaussian distribution (\ref{2.12}), 
the average can be easily performed. 
Since the impact of the 
metric fluctuation on the field is expressed by
 $w_0 = \mu_0 /{\sqrt {1 + \omega^2}}$
instead of the origin amplitude $\mu_0$,
we introduce the new distribution function
\be \n{3.18}
\rho_\omega(w_0) = {1 \over 2 \pi \sigma_\omega^2} 
\exp (-{w_0^2 \over 2 \sigma_\omega^2})\, ,
\ee
where
\be \n{3.19}
 \sigma_{\omega} = {\tilde{\sigma}_\omega \over {\sqrt {1+ \omega^2}}}\, .
\ee
It is easy to check that 
\be \n{3.20}
2\pi\, \int_{0}^{\infty}\, dw_0\, w_0\rho_\omega(w_0)=1\,\, , \hspace{0.5cm} 
<w_0^2>_{\omega} = 2 \sigma_\omega^2\,.
\ee
where the average is now taken over $w_0$ with the distribution
function $\rho_{\omega}(w_0)$.

Using using (\ref{3.9}) and (\ref{3.18}), the average of $\bar \Delta^+$ gives
\be\n{3.20b}
\av{\Delta^+(u|w)}= {1 \over \sigma_{\omega} \sqrt{2 \pi}}\,\, 
\exp \left[-{(e^{-u} - w)^2\over
2 \sigma^2_{\omega}} \right]\, .
\ee
Thus the Gaussian character of $\rho_\omega$, the distribution
of metric fluctuations, is preserved and now 
characterizes the distribution of light rays. 
The same result applies to backward propagation since one has
$\av{{}^{-\!\!}\Delta (w|u)} = e^{-u}\av{\Delta^+ (u,w)}$. 

At fixed $u$, $\av{\Delta^+}$ gives the image on ${\cal J}^-$ of the
given $u$-ray sent from ${\cal J}^+$. This image  gives the probability
that this ray  reaches the interval $dw$ centered on $w$.     For $u\ =
\infty$, $\av{\Delta^+}$ therefore gives the image of  the event
horizon. Since we get a Gaussian distribution centered at $w=0$, the
position corresponding to the horizon in the unperturbed geometry, this
indicates that the horizon is {\em smeared} by the stochastic
fluctuations.  And eq. (\ref{3.20b}) shows that the global properties 
of the field propagation are sensitive to this smearing of the event
horizon.

\subsection{Scattering operator  in a fluctuating black-hole geometry}

The relations (\ref{3.5}), (\ref{3.9}) and (\ref{3.20b}) solve the classical
scattering problem in our fluctuating geometry. We now demonstrate how
this problem can be solved in a more formal way. This more formal
approach will be very useful  for understanding the modifications of
the propagation induced by the metric fluctuations.

In order to simplify the algebra it is most convenient to introduce 
a new coordinate on ${\cal J}^+$
\be\n{3.21}
y=e^{-u}\, .
\ee
The reason is the following. In terms of $y$, eq. (\ref{2.22}) becomes
\be\n{3.22}
w=W_\phi(y)\equiv y+w_0 \sin(\phi +\phi_0)\, .
\ee
Therefore, the effect of the metric fluctuations is simply to shift $y$
with respect to $w$. The image  ${\Phi}^+$ of a wavepacket on ${\cal
J}^+$ can be considered as a function of $u$ or of $y$.  In order to
avoid confusion, we shall keep the notation ${\Phi}^+(u)$ for the
function of $u$ and shall use the notation ${\mathbf \Phi}^+(y)$
whenever the image is considered as a function of $y$.

In terms of $y$, $\bar{\Delta}^+(u|w)$
given by (\ref{3.9}) takes the form
\be
\bar{\mathbf \Delta}^+(y |w)={1\over \pi } {\vartheta(w_0^2-(y-w)^2)
\over \sqrt{w_0^2-(y-w)^2}}
\, ,
\n{3.23}
\ee
and relation (\ref{3.6}) becomes
\be
\bar{\mathbf \Phi}^+(y) = \frac{1}{\pi}\,\int_{-w_0+y}^{w_0+y}\frac{dw\,
{\Phi}^-(w)}{\sqrt{w_0^2-(y-w)^2}}\,. \n{3.24}
\ee
According to our definition (\ref{3.21}) the coordinate $y$ takes only
positive values. But we can  use relation (\ref{3.24}) to define
$\bar{\mathbf \Phi}^+(y)$ (at least formally) as a function on the
complete $y$ axis\footnote{The geometrical meaning of negative 
values of $y$ will be presented in the next subsection.}.
In this case, eq. (\ref{3.24})
defines a linear operator acting on the space of functions 
defined on the real axis. We call this operator 
$\bar{{\bf {\sf D}}}$. Thus  (\ref{3.24}) 
takes the form
\be
\bar{\mathbf \Phi}^+ = \bar{{\bf {\sf D}}}\,\, \Phi^-\,. \n{3.25}
\ee

A formal representation for this operator can be easily obtained 
by exploiting the fact that eq. (\ref{3.22}) is a linear map.
Indeed, for any realization of the geometry we have
\be\n{3.26}
{\mathbf \Phi}^{+}_{\phi}(y)={\Phi}^-(y+w_0 \sin(\phi +\phi_0))
=e^{w_0 \sin(\phi +\phi_0)\partial}\, \Phi^-(y)\, .
\ee
Here $\partial$ is the operator of differentiation with respect of
the argument of the function. This relation can be rewritten as
\be
{\mathbf \Phi}^+_{\phi}= {{\bf {\sf D}}}_{\phi}\,\, \Phi^-\,, \n{3.26a}
\ee
where
\be\n{3.26b}
{\bf {\sf D}}_{\phi}=e^{w_0 \sin(\phi +\phi_0)\partial}\, .
\ee
Notice that this shift operator bears some similarities with that
introduced in \cite{thoooft}. 

Let us first perform the average over the phase. Using the integral
representation of the Bessel function of zero index
\be\n{3.27}
J_0(b)={1\over \pi}\int_0^\pi d\phi \,\, e^{ib\cos\phi}\, ,
\ee
we get
\be
\n{3.28}
\bar{{\bf {\sf D}}}= \int_0^{2 \pi} {d\phi \over 2 \pi} {\bf {\sf D}}_{\phi} =
J_0(-iw_0\partial)\, .
\ee

Let us now show that (\ref{3.23})
is a direct consequence of (\ref{3.28}). For this purpose it is
convenient to adopt the Dirac notations and to write the 
functions $\bar{\mathbf \Phi}^{+}$ and
$\Phi^-$ as ket-vectors $|\bar{\mathbf \Phi}^{+}\rangle$ and
$|\Phi^{-}\rangle$, respectively. Then, in the ``coordinate'' representation,
we have
\be\n{3.29}
\bar{\mathbf \Phi}^{+}(x)=\langle x|\bar{\mathbf \Phi}^{+} \rangle 
\, ,\hspace{0.5cm}
\Phi^{-}(x)=\langle x|\Phi^{-} \rangle \, ,
\ee
and equation (\ref{3.25}) 
takes the form
\be
\bar{\mathbf \Phi}^+ (x) = \int_{-\infty}^{\infty} dx' 
\langle x|\bar{{\bf {\sf D}}}|x'  \rangle \,
 \Phi^-(x') \,. \n{3.30}
\ee
Relation (\ref{3.28}) implies that the operator $\bar{\bf {\sf D}}$ is
diagonal in the ``$p$-representation''. 
This directly follows from the linearity of eq. (\ref{3.22}).
Explicitly one has
\be\n{3.31}
\langle p|\bar{\bf {\sf D}}|p'  \rangle = \delta(p-p')\, J_0(w_0
p)\, .
\ee
Then, using
\be\n{3.32}
\langle p| x \rangle = {1\over \sqrt{2\pi}} \exp{(ipx)}\, ,
\ee
we get 
\be\n{3.33}
\langle x|\bar{\bf {\sf D}}|x'  \rangle \, =
\int_{-\infty}^{\infty} dp \int_{-\infty}^{\infty} dp'
\langle x| p \rangle \langle p|\bar{\bf {\sf D}}|p'  \rangle 
\langle p'| x' \rangle =
{1\over 2\pi}\int_{-\infty}^{\infty} \, \, dp \,
e^{-i(x-x')p}\, J_0(w_0 p)\, .
\ee
Calculating the integral (see e.g. \cite{GrRy:94}) we get
\be\n{3.34}
\langle x|\bar{\bf {\sf D}}|x'  \rangle 
= {1\over \pi}{\vartheta(w_0^2-(x-x')^2)\over \sqrt{w_0^2-(x-x')^2}}\,
.
\ee
Thus  $\langle y|\bar{\bf {\sf D}}|w  \rangle$
equals $\bar{{\mathbf \Delta}}^+ (y \vert w)$ given by (\ref{3.23}).

Up to now we have only been dealing with the stochasticity
connected with the phase $\phi$. Let us discuss what
happens when we average the operator ${{\bar{\bf \sf D}}}$ over
the amplitude of the metric fluctuations. We again assume that the 
probability distribution is Gaussian (\ref{3.18}). Using the relation
(see \cite{Wats:66} p.393, eq. 13.3.1)
\be\n{3.35}
\int_0^{\infty} dz z e^{-z^2} J_0(\alpha z)={1\over 2}e^{-\alpha^2/4}\, ,
\ee
we get that the average of $\bar{\bf {\sf D}}$
over the fluctuation amplitude is 
\be\n{3.36}
\av{\bf {\sf D}} = 
\exp\left(-{{\sigma}^2_{\omega}\partial^2\over 2}\right)\, .
\ee
This simple form results from the linearity of the map
(\ref{3.22}) and the Gaussian character of $\av{\Delta^+(u|w)}$ (\ref{3.20b}).

The generalization of these results to a spectrum of
metric fluctuations with different frequencies is straightforward.
Indeed, since  $\av{\bf {\sf D}}$ has a simple exponential form, 
it enjoys the factorization property (\ref{2.11}). Thus,
using  (\ref{2.17}), its total ensemble average 
\be\n{3.37}
\ll \!{\bf {\sf D}}\!\gg = 
\exp\left(-{{\sigma}^2_{\ind{eff}}\partial^2\over 2}\right)\, ,
\ee
where
\be\n{3.38}
{\sigma}^2_{\ind{eff}}=\sum_{\omega}\, {\sigma}^2_{\omega}\, .
\ee
Eq. (\ref{3.38}) shows that the effect of the whole spectrum of metric
fluctuations is to give rise to a single length which weights higher
order derivative terms. This shows that the details of the fluctuations
spectrum play no significant role for the scattering operator $\ll \!{\bf
{\sf D}}\!\gg$.

In resume, finding the image $\Phi^+$ on ${\cal J}^+$ of a wavepacket
determined by its data $\Phi^-$ on ${\cal J}^-$ is a ``classical'' ${\bf {\sf
S}}$-matrix problem. In the absence of metric fluctuations,
 the linear operator ${\bf {\sf S}}_0$ 
relating $\Phi^-$ and ${\mathbf \Phi}^+$ is trivial (equal to 1)
if we use the coordinates $y$ and  $w$. In
the presence of a given realization of the metric the operator is
given by eq. (\ref{3.26b}).  When one computes the ``mean''
propagation by considering a stochastic ensemble of metric
fluctuations,  this simple relation is modified and takes the form 
(\ref{3.28}), (\ref{3.36}) or (\ref{3.37}) according to the ensemble
of metric fluctuations that one considers.

The extremely simple form of the operators ${\bf {\sf D}}$ in the
``p''-representation allows one to make a few general
observations. In particular, we have
\be\n{3.39}
\ll \!{\mathbf \Phi}^{+}(p)\!\gg \equiv \langle p|
\ll \!{\mathbf \Phi}^{+}\!\gg \rangle=
e^{-{\sigma}^2_{\ind{eff}}\,p^2/2}\,{\Phi}^{-}(p)\, .
\ee
Thus only the high frequency components  (i.e. of the order of
${\sigma}_{\ind{eff}}^{-1}$ and greater) of the initial wave packet 
$\Phi^-$ are strongly affected by the fluctuations of the geometry. 
Therefore  the forward propagation of any smooth (i.e. of Fourier
content much below ${\sigma}_{\ind{eff}}^{-1}$)   wave packet defined
on ${\cal J}^-$ will not be significantly affected by the metric
fluctuations.  In other words, for classical black hole physics, the
metric fluctuations are irrelevant if, as indicated in \cite{York:83,
CEMNP}, their  ``mean'' amplitude ${\sigma}_{\ind{eff}}$ is of the
order of the Planck length or smaller than it.

\subsection{Backward in time scattering}

Eventhough eq. (\ref{3.22}) is perfectly symmetric in $w$ and $y$,
backward propagation is dramatically affected by the metric 
fluctuations. 

To settle the discussion, we first clarify the geometrical  meaning
of negative $y$. To ease this analysis, we start with backward
propagation in the absence of fluctuations.  In this case, for
positive $y$, eq. (\ref{3.22}) gives $w=y$. However, since
${\cal{J}}^+$ is not a complete Cauchy surface, we need to consider
the union of ${\cal{J}}^+$ and the whole event horizon $u= \infty$ in
order to have a complete  Cauchy surface. Thus we must introduce a
coordinate along the horizon.  The simplest choice consists in
considering the negative values of $y$ defined again by $y=w$. Indeed
the negative $y$ axis so defined  covers the horizon from $r=0, w=0$
inside the collapsing shell till $w=-\infty$.  Thus,  the real axis $y
\in (-\infty, \infty)$ forms a  complete Cauchy surface and  the
functions ${\mathbf \Phi}^+(y)$  determine their image $\Phi^-(w)$ on
${\cal J}^-$ for all $w$. 

This procedure also applies for any given realization of the
fluctuating geometry. Indeed, the event horizon, when continued 
backward for negative $v$ in the inside flat geometry (\ref{2.3}),
reaches $r=0$ at advanced $w$ time 
$W_\phi(y=0)= w_0 \sin(\phi +\phi_0)$. 
Therefore  the negative half line $y\in (-\infty, 0)$ defined by eq.
(\ref{3.22}) still covers the whole horizon and  $y\in (-\infty,
\infty)$ forms a complete final Cauchy surface. Since this is valid for
any realization, it is meaningful to use the coordinate $y\in
(-\infty, \infty)$ after having averaged over $\phi$. 

For $y>0$, ${\mathbf \Phi}^+(y)$ gives the
value on ${\cal J}^+$,  while for $y<0$ it gives the value on the
horizon.  For regular  ${\mathbf \Phi}^+(y)$,
in virtue of the symmetrical role played by $y$ and $w$  in eq.
(\ref{3.22}), the averaged image on ${\cal J}^-$ is determined by
the same scattering operator $ {\bf {\sf D}}$ which governed
forward propagation. In the case of the full ensemble average, 
one has  
\be
\ll \!{ \Phi}^-\!\gg  = \ll \!{\bf {\sf D}}\!\gg \, 
{\mathbf \Phi}^+\,. \n{3.39b}
\ee
In Fourier transform with respect to $w$ and $y$
this gives 
\be\n{3.40}
\ll \!{\Phi}^-(p)\!\gg = \exp{(-\sigma_{\ind{eff}}^2 p^2/2)}\, 
{\mathbf \Phi}^+(p)\, .
\ee

Of special interest for studying Hawking radiation are the final images such
that  no incoming field emerges from the horizon for any realization
of the geometry. They are of the simple form
\be
{\mathbf \Phi}_{out}^+(y)= \vartheta(y) {\mathbf \Phi}^+(y) \, .
\label{3.38b}
\ee
Unless ${\mathbf \Phi}^+(y)$ vanish sufficiently rapidly
when $y \to 0$, these modes are singular at $y=0$. 

This problem will be studied in detail in Section 5.
It reveals the important role played by
the fluctuating horizon geometry for backward propagation. The
asymmetry between backward and forward propagation comes  from the
fact that the inertial time on ${\cal J}^+$ which characterizes 
out-frequencies $\lambda = \partial_u$  is $u$ and not $y= e^{-u}$.
Then, the so defined out-frequencies are exponentially blue-shifted
when  propagated backward near the event horizon. This purely
kinematical effect is at the origin of the trans-Planckian problem
and  has here dramatic consequences since higher derivative terms are
present. Indeed, eq. (\ref{3.40}) when applied to out-functions
(\ref{3.38b}) which vanish for negative $y$ gives, in the position
representation,
\be
\ll \!{\Phi}^-(w)\!\gg 
= \left[\exp\left(- {1\over 2}(\sigma_{\ind{eff}} \, e^{u}\partial_u )^2\right)\,
\, {\Phi}^+(u)\right]_{u=-\ln w}
\, .
\n{3.41}
\ee
The dramatic consequences can now be seen: how ever smooth is the
final data ${\Phi}^+(u)$, the fluctuations of the geometry will
inevitably affect its backward propagation if it is centered around a
sufficiently late retarded time. Moreover if it does not vanish
sufficiently fast  (i.e. faster than\footnote{This condition on the
decrease of the wave packet already appeared in the
literature \cite{MP952} in the Unruh detector context (see also
\cite{MP95} for its transposition in black hole physics). In the
Unruh  case, if one imposes that the fluxes of energy emitted by the
accelerated  detector be regular in an inertial frame, the coupling
between this detector and the field must decrease faster than $e^{-a
\tau}$ where $a$ is the acceleration and  $\tau$ the detector's
proper time.} $e^{-u}$) when one approaches the horizon, the
asymptotic behaviour of $\ll \!{\bf {\sf D}}\!\gg$ intervenes.

We shall further analyze these points in quantum mechanical terms in
Section 5. Before presenting this material we shall show that the
asymptotic properties of  Hawking radiation are not significantly
affected by  the fluctuations when their average amplitude
$\sigma_{\ind{eff}}$  is much smaller than the Schwarzschild radius
($=1/2$ in our units).

\section{Hawking Radiation} \setcounter{equation}0

\subsection{Green function in the $in$-vacuum}

The simplest way to understand why the metric fluctuations do not
significantly modify the asymptotic properties of Hawking radiation
consists in analysing the Green function (more specifically, the
positive frequency Wightman function\cite{BiDa:82}) evaluated in the initial
vacuum state. Indeed, as shown in \cite{HAA},  when the short distance
expansion of this  function evaluated near the event horizon  reduces
to the standard (Hadamard) behavior, Hawking radiation obtains 
on ${\cal{J}}^+$.

In the absence of fluctuation, the $in$-vacuum is the vacuum state
with respect to positive frequencies defined on ${\cal{J}}^-$ in
terms of  the inertial advanced time derivative $i\partial_v$. 
Therefore, for late time, the $u$-part of the  Green
function evaluated in this vacuum is controled by $V(u)= -1 -
e^{-u}$. Explicitely, one has
\ba\n{4.1}
G^{in}(u, u') &=& \int_0^\infty {dp \over 4 \pi p} e^{i p(V(u) - V(u') 
+ i\epsilon)}
\nonumber\\
&=&{1 \over 4\pi} \ln( V(u) - V(u') + i\epsilon)
= {1 \over 4\pi} \ln( - e^{-u} + e^{-u'} + i\epsilon) \, .
\ea
The thermal and  steady character of Hawking radiation follows  from
the exponential relation between $V$ and $u$, which remains unchanged
for arbitrary large time $u$, and from the fact that for small $V$
intervals ($\delta V \ll 1$), the $in$-Green function behaves like
$\ln\delta V$.
Therefore these are the two conditions which should be met in order to
see if the metric fluctuations lead to modifications of Hawking
radiation.

In a given realisation of the metric fluctuation,  one simply
replaces the unperturbed relation $V(u)=-1 - e^{-u} $ by its
modified version eq. (\ref{2.20}). Upon considering the ensemble of
geometries, one averages the Green function over the ensemble.
After averaging over phases, the mean function is thus given by
\ba
\bar{G}^{in}(u, u') &=& \int_0^{2 \pi} {d\phi \over 2 \pi}
G^{in}_\phi(u, u')
\nonumber\\
 &=&{1 \over 8\pi^2} \int_0^{2 \pi} d\phi 
\ln( V_\phi(u) - V_\phi(u') + i\epsilon)\, .
\n{4.2}
\ea
First, we notice that when using the simplified expression (\ref{2.22})
Hawking radiation is not at all affected since the constant shift cancels
out. The reason is clear: in each realization, two nearby points are
shifted by almost the same amount, see Figure 2.

Therefore only higher order effects might affect Hawking radiation.
To compute them, it is again usefull to use $y=\exp(-u)$.
In terms of $y$, using eq. (\ref{2.20}), one has
\be
V_\phi(u) - V_\phi(u')=F_\phi(y')-F_\phi(y) + O(w_0^2)\, ,
\n{4.3}
\ee
where
\be\n{4.4}
F_\phi(y)=y(1+a\cos(\phi+2\phi_0)+b\sin(\phi+\psi_0 -\omega  \ln y))\, .
\ee
Here $\psi_0=\phi_0-\varphi_{\Gamma}$. The explicit form of the coefficients
$a$ and $b$ can be easily obtained from eq. (\ref{2.20}).
They are not important for us now. It is
sufficient to mention that both these coefficients are of
first order in $w_0$. For close points $y'=x+z/2$
and $y=x-z/2$ ($z\ll 1$) one has
\be
F_\phi(y')-F_\phi(y) =z\left[1+\sum_{n=0}^{\infty} c_n(\phi)
 z^{2n}\right]\, .
\n{4.5}
\ee
All the coefficients $c_n$ are of the first order in $w_0$ since
$F_\phi$ has been linearized.
In particular, the first coefficient  
\be\n{4.6}
c_0(\phi)=a\cos(\phi+2\phi_0)+b\sin(\phi+\psi_0 -\omega  \ln
x)- b\omega \cos(\phi+\psi_0 -\omega  \ln x)\,.
\ee
is a linear superposition of sine and cosine
whose arguments are linear in $\phi$. It is easy to check that this
is also true for all other coefficients $c_n(\phi)$. Then, using
(\ref{4.3}) and (\ref{4.5}), we can  expand $\ln( V_\phi(u) -
V_\phi(u') + i\epsilon)$ in powers of $w_0$ and conclude that
after averaging over $\phi$ one has
\be\n{4.7}
\bar G^{in}(u, u')={G^{in}(u, u')}+w_0^2 H(u, u')\, ,
\ee
since the averaged value of all first order terms in $w_0$ vanishes.
The crucial point is that $H(u, u')$ is finite when $u \to u'$.  This
implies that the corrections to Hawking radiation are at least of
second order in $w_0$ and non-diverging in the late time regime.  Upon
considering the average over the fluctuation amplitudes the same result
holds. These points are  explicitely verified  in the next section.

\subsection{Quadratic corrections to Hawking Flux}

The (quantum mechanical) mean energy flux of  Hawking radiation
measured at ${\cal J}^+$,   $dE/du$, was obtained in \cite{BaFrPa:99} 
in the case of the oscillating Vaidya (\ref{2.8}) metric with the phase
$\phi$ in eq. (\ref{2.18}) put equal to zero.  The corrections were
computed up to second order in the fluctuation amplitude $\mu_0$ by
using the 2D model based on eq. (\ref{3.2}). In that case, the energy
flux  can be decomposed into the sum of a permanent part
$(dE/du)^{\ind{perm}}$ and a fluctuating part $(dE/du)^{\ind{fluct}}$
\be
dE/du = (dE/du)^{\ind{perm}} +  (dE/du)^{\ind{fluct}}\,. \n{4.8}
\ee
The permanent part is equal to
\be
 (dE/du)^{\ind{perm}} = \frac{\kappa ^2}{48 \pi} \left[1 + 
\frac{1}{2}\, w_0^2 (1+\omega^2)q^2(\omega) + w_0^2 \right]\,, \n{4.9}
\ee
where $\kappa =(8\pi M)^{-1}$ is the surface gravity of the unperturbed
black hole, and
\be
q(\omega) = \frac{\sqrt{2\pi}}
{\sqrt{\omega (e^{2\pi\omega}-1)}}\,.
\n{4.10}
\ee
In the absence of metric fluctuations, the permanent part of
the flux reduces to its usual value,
$\kappa ^2 /48 \pi$. The corrections due to these fluctuations are
second order  in $w_0$. The fluctuating part  $(dE/du)^{\ind{fluct}}$ is a
linear combination of terms oscillating with frequency $\omega$ and
$2\omega$  and, by definition, vanishes when integrated over the
retarded time $u$. 

We shall now compute this flux in a stochastic metric 
by averaging its value over the phase
and amplitude of the metric fluctuations.
We first study the influence of a single stochastic phase $\phi$.
In particular we shall check that averaging
over this phase {\it and} over $u$ does not affect the
value of permanent part of the flux.
For this purpose, as in \cite{BaFrPa:99},
we calculate $dE/du$ from the 2D expression 
\be
{dE \over du}(u, \phi)= \frac{\kappa ^2}{12 \pi}
 (dV_\phi /du)^{1/2} \frac{d^2}{du^2}
\left[(dV_\phi/du)^{-1/2} \right]\,, \n{4.11}
\ee
where $V_\phi(u)$ now depends on the phase $\phi$, see (\ref{2.20}). 
Since we want to compute the quadratic corrections  in $w_0$, we need
to  know $V_\phi(u)$ up to second order in $w_0$. In the late time regime
and to this order in $w_0$, $V_\phi(u)$ can be decomposed
according to the dependence of the various terms in $u$.
Using eqs. (4.51-60) in \cite{BaFrPa:99}, one gets
\be
  V_\phi(u) = V_0 (\phi) -e^{-u} \left[  V_1 (\phi)
   + V_2 (u,\phi) - {w_0^2 \over 2} u \right]\,. \n{4.12}
\ee
By definition $V_0$ and $V_1$ do not depend on $u$. Note that $V_0$,
$V_1$ and $V_2$ parametrically depend on the frequency $\omega$ of the
fluctuations.

For a given $\phi$, i.e. for a given realization of the geometry, one
easily sees from (\ref{4.11}) that $dE/du$ does not depend on the
values of $V_0$ and $V_1$. Therefore one is free to put $V_0 = 0$ and
$V_1 = 1$ and to work with $V_\phi(u)$ as 
\be
 V_\phi(u) =  -e^{-u} \left[ 1
   + V_2 (u,\phi)  - {w_0^2 \over 2}u \right]\,. \n{4.14}
\ee
The novelty with respect to eq. (\ref{2.20}) arises from the
fact that $V_{\phi} (u)$ is now developed up to second 
order in $w_0$. As $V_2 (u,\phi)$ has no $0$-th order term
in $w_0$ it can be decomposed as
\be
 V_2 (u,\phi) = w_0  V_{2,1}(u,\phi) + 
   w_0^2  V_{2,2}(u,\phi)\,. \n{4.15}
\ee 
The first term is already known, see eq. (\ref{2.20}), 
\be
V_{2,1}(u,\phi) = q(\omega)\, 
\sin (\phi + \phi_0 + \omega u - \varphi_{\Gamma} )\,, \n{4.16}
\ee
where $\varphi_{\Gamma}(\omega)$ is a phase independent of $u$.
$V_{2,2}(u,\phi)$ is a complicated expression containing oscillating
terms with respect to $u$ and $\phi$. It is of little interest to write it explicitely.
In what follows we shall only need its averaged value with respect to
$\phi$ since we are interested by the expression of the 
averaged energy flux to second order in $w_0$.

In terms of these two functions, the energy flux reads
\ba
\frac{dE}{du}(u,\phi) &=& \frac{\kappa ^2}{48 \pi}
\left[ 1 + w_0^2 -2w_0 (V_{2,1}' - V_{2,1}'')
+ 2w_0^2 (V_{2,1} - V_{2,1}')(V_{2,1}' - V_{2,1}''')  \right.
\nonumber\\
&& + \quad \left. 
3w_0^2 (V_{2,1}' - V_{2,1}'')^2 
-  2w_0^2 (V_{2,2}' - V_{2,2}''') \right]\,, \n{4.17}
\ea
where $'$ stands for $\partial / \partial {u}$.

Using as in Section 2 a ``bar'' to denote averaging over $\phi$
one gets
\be
\overline{\frac{dE}{du}} (u) =  \frac{\kappa ^2}{48 \pi}
\left[ 1 + w_0^2 + \frac{1}{2}\,w_0^2(1+\omega^2) q^2(\omega)
- 2w_0^2 \overline{(V_{2,2}' - V_{2,2}''')}\right]\,. \n{4.18}
\ee
The first $w_0^2$ term was already present in eq. (\ref{4.9})
and is due to the last term in  (\ref{4.12}). 
The second one comes from quadratic terms in 
$V_{2,1}$ given by (\ref{4.16}). The last term depends on $u$ and
is equal to
\be
\overline{V_{2,2}' - V_{2,2}'''} = (1-2\omega^2)q(\omega) \cos(\omega u - 
\varphi_{\Gamma}) + \omega (3-4\omega) \sin(\omega u - 
\varphi_{\Gamma})\,. \n{4.19}
\ee
These results show that after taking the averages over $\phi$
and over $u$ one gets back the same permanent contribution to the
Hawking mean flux as in a given realisation of 
the geometry, see eq. (\ref{4.8}). 
The reason is that averaging over $u$ erases all dependence in $\phi$.

Eq. (\ref{4.18}) can be easily extended to a spectrum
of metric fluctuations according to the lines presented in section 2.2. 
Limiting ourselves to the permanent part of  (\ref{4.18}) we obtain
\be\n{4.20}
\left( \overline{\frac{dE}{du}} \right)^{\ind{perm}} = \frac{\kappa ^2}{48 \pi}
\left[ 1 + w_0^2 \left(1+ \frac{1}{2}\,(1+\omega^2) q^2(\omega) \right) \,\right]\, ,
\ee
which gives the average over the phase of a single
fluctuating mode of frequency $\omega$.
When using the Gaussian distribution $\rho_{\omega}$, see  
(\ref{3.18}), the average over the amplitude $w_0(\omega)$ of the 
fluctuating mode gives
\be\n{4.21}
<\left( \frac{dE}{du} \right)^{\ind{perm}}\!\!>_{\omega} = 
\frac{\kappa ^2}{48 \pi}
\left[ 1 + 2\sigma_{\omega}^2\left(1+ \frac{1}{2}\,(1+\omega^2) q^2(\omega) \right)
\, \right]\, ,
\ee
where  (\ref{3.20}) has been used.

Upon considering the whole spectrum of fluctuations, we get
\be\n{4.22}
\ll\!\left( {dE \over du} \right)^{\ind{perm}}\!\gg = 
\frac{\kappa ^2}{48 \pi}\left[ 1 + 2 \sum_\omega 
\sigma_{\omega}^2
\left(1+ \frac{1}{2}\,(1+\omega^2) q^2(\omega) \right)
\, \right]\, .
\ee
To perform this last summation requires the 
knowledge of the spectrum, here represented
by the set of $\sigma_\omega$.

\section{Fluctuating Geometry and Trans-Planckian Problem}

\subsection{Hawking radiation for modified dispersion relations}
In this Section we shall establish the close analogies between the 
 backward propagation in a fluctuating metric and the altered propagations
which have been recently studied and which result from the 
modifications of the dispersion relation in the high frequency regime.

Before presenting the technical details, it is appropriate to recall
the following points.  These models have been introduced in order to
show that the mutilation of the dispersion relation for frequencies
higher than $\omega_c$, which is the equivalent of $\sigma_{\ind{eff}}^{-1}$ in our
case, in no way affect the (low energy) properties of Hawking
radiation, namely stationarity and thermality. Following the original
work of Unruh\cite{Unruh81, Unruh95}, many models have been analysed. Their
common property is that the Dalembertian is modified by the
addition of higher derivative terms weighted by negative powers of
$\omega_c$.   In these works, the modifications have been inspired by
hydrodynamics\cite{Visser}, electrodynamics in a dielectric medium\cite{Reznik}, 
field theory on a lattice theory\cite{Jac99}, string theory\cite{BS2s} or by
guessing what the physics near a horizon might be\cite{BMPS95}. In all
these models, the following properties obtain
\begin{enumerate}
\item Fowardly propagated wave packets are unaffected by the
modification of the dispersion relation  as long as their
in-frequency content is much below the  critical frequency
$\omega_c$. 
\item No significant modifications of the asymptotic properties of
Hawking radiation as long as the surface gravity satisfies $\kappa \ll
\omega_c$.
\item Dramatic modifications of backward propagated late-time wave packets of
out-frequency $\lambda$  when the blue shifted value $\lambda e^u$
reaches $\omega_c$.
\end{enumerate}
It should be already clear to the reader that our effective propagation 
in a fluctuating metric possesses many similarities with 
these models. Our aim is now to establish  the parallelism in 
simple and analytical terms. To this end we shall exploit the 
following facts.

First we exploit the stationarity of the unperturbed background
geometry by considering the backward propagation  of a plane wave of
out-frequency $\lambda$ defined on ${\cal{J}}^+$. We shall consider its
image on $v=0$ rather than on ${\cal{J}}^-$ in order to ignore the
backward propagation from $v=0$ till ${\cal{J}}^-$ which is very much 
dependent on a model of a collapsing body  and presumably irrelevant
for black hole physics. The relationship with what we did in the former
Section is straightforward since $w$, defined on ${\cal{J}}^-$,
corresponds to  $2r-1$ on $v=0$ in the dimensionless coordinates
defined in Section 1 where $\kappa =1$. 
 
Secondly, we work in Fourier transform with respect to $w$
as it provides an elegant characterization of in-vacuum. 
Since $r$ is an afine parameter
along $v=\mbox{const}$, $\,-i\partial_r$ corresponds to 
positive frequencies measured by inertial observers when
they cross the event horizon. Therefore, when infalling observers
see no particles, it means that the state
of the radiation field corresponds to vacuum with respect to 
frequency modes of positive $-i\partial_r$. 
Moreover, since $\partial_r \vert_{v=0}= 2 \partial_w\vert_{\cal{J}^-}$,
the initial vacuum with respect to $w$ coincides with vacuum as seen
by infalling observers, see \cite{GO, Jac93}.

Lets now review how these concepts translate in mathematical 
terms and how they can be used to compute the consequences
of modifying the dispersion relation at high frequencies.
In this we shall present the ``alternative'' model
of \cite{BMPS95} for its simplicity and its generality.

The 2D Dalembertian (\ref{3.2}) for out-going modes with out-energy
$\lambda$ propagating near the event horizon ($r -1/2 \ll 1/2$) 
gives, in the unperturbed Vaydia metric (\ref{2.2}),
\be
( 1 - 2r) i \partial_r \varphi_\lambda = 2 i \partial_v  \varphi_\lambda 
= 2 \lambda  \varphi_\lambda \, .
\ee
Along $v=0$, it is useful to express $r$ in terms of 
$w=2r-1$. This leads to the simplified equation
\be
w \partial_w  \varphi_\lambda =  i\lambda  \varphi_\lambda \, .
\label{r11}
\ee
The general solution is of the form $\varphi_\lambda = A \vartheta(w)
w^{i \lambda} + B \vartheta(-w)(-w)^{i \lambda}$. 
The normalized out-mode describing the one particle state
of energy $\lambda$ is given by 
\be
\varphi_\lambda^{out}
=  \vartheta(w) {w^{i \lambda} \over \sqrt{4 \pi \lambda}}\,.
\label{r15}
\ee
On the other hand, the mode leaving on the other side
of the horizon describes the partner of this Hawking quantum.

In Fourier transform $p = i \partial_w$ eq. (\ref{r11}) becomes
\be
 \partial_p  
\left( p \varphi_\lambda \right)
= - i\lambda  \varphi_\lambda \, .
\label{r13}
\ee
In terms of $p$,  the normalized in-mode $\varphi_\lambda^{in}$
contains  only positive $p$ and is thus given by 
$\varphi_\lambda^{in} = \vartheta(p) p^{-i\lambda -1} /{ \sqrt{4 \pi
\lambda}}$. To obtain the Bogoliubov coefficients encoding the
thermal flux of outgoing quanta it suffices to inverse Fourier 
transform this
mode and use the definition (\ref{r15}). Explicitely one has
\ba
\varphi_\lambda^{in}(w) &=&
\int^\infty_0 {dp \over {\sqrt{2 \pi}}} e^{ip w } \varphi_\lambda^{in}(p) 
\nonumber\\
&=& 
{\Gamma(-i\lambda)  \over  {\sqrt{2 \pi}}}
{(\epsilon - i w)^{i\lambda} \over  \sqrt{4 \pi \lambda}}
\nonumber\\
&=& {\Gamma(-i\lambda)  \over  {\sqrt{2 \pi}}}
\left[ e^{\pi \lambda /2} \varphi_\lambda^{out} +
 e^{-\pi \lambda /2} \vartheta(-w) {(-w)^{i \lambda} \over  \sqrt{4 \pi \lambda}}
\right]\, ,
\label{r43}
\ea
where $\Gamma(z)$ is the Euler gamma function
and where $\epsilon$ is small and positive. It fixes 
the relative weights of $w^{i \lambda}$ for 
positive and negative real values\cite{GO}, as shown  
explicitely in the third line.
This relative weight determines in turn the ratio
of the Bogoliubov coefficients $\alpha_\lambda$
and $\beta_\lambda$: 
$\vert \alpha_\lambda / \beta_\lambda \vert
= e^{\pi \lambda}$.

For later purpose, we notice that the same result can also be obtained
the other way around, by Fourier transforming the out-mode (\ref{r15}).
In this case one gets
\ba
\varphi_\lambda^{out}(p) &=&
\int^\infty_0 {dw \over {\sqrt{2 \pi}}} e^{-ip w } \varphi_\lambda^{out}(w) 
\nonumber\\
&=& 
{\Gamma(i\lambda+1)  \over  {\sqrt{2 \pi}}}
{(\epsilon + i p)^{-i\lambda-1} \over  \sqrt{4 \pi \lambda}}
\nonumber\\
&=& { (-i) \Gamma(i\lambda+1)  \over  {\sqrt{2 \pi}}}
\left[ e^{\pi \lambda /2} \varphi_\lambda^{in} -
 e^{-\pi \lambda /2} \vartheta(-p) {(-p)^{-i\lambda -1} \over 
  \sqrt{4 \pi \lambda}} 
\right] \, .
\label{r432}
\ea
The only subtility concerns again the prescription given by
$\epsilon$. It arises this time from the fact that  the final wave
packet vanishes on the other side of the  horizon. Upon Fourier
transform, this fixes the  relative weights on the positive and
negative real $p$ axis. This can be used again to determine  the
ratio of the Bogoliubov coefficients.

We are now in position to modify the dispersion
relation. In a flat metric, this relation is simply $p = \lambda$.
To modify it, we write it as $g(p)= \lambda$. The only condition that 
$g$ must satisfy is that for small $p$ is behaves as 
$g(p) = p (1 + O(p/ \omega_c))$. 
Near the horizon, for $\omega_c \gg 1$ and up to a normal
ordering ambiguity of $\partial_p$ and $g(p)$
which plays no role in a WKB approximation, 
eq. (\ref{r13}) becomes
\be
 \partial_p \left( g(p) \tilde \varphi_\lambda \right)= - i\lambda \tilde
\varphi_\lambda \,.
\ee
The general solution of this modified  equation is given by
\be
\tilde \varphi_\lambda = A \vartheta(p){ e^{-i \lambda \int^p dp'/g(p')} 
\over g(p) }
+ B  \vartheta(-p){ e^{-i \lambda \int^p dp'/g(p')} \over g(p) } \, .
\label{r12}
\ee

In these terms, the three properties listed above are easily 
obtained. The first one follows from the definition of $g(p)$ which
deviates from linearity only for $p > \omega_c$. The second point is
verified by using the fact that  in-modes characterizing vacuum for
infalling  observers still contain only positive $p$. The proof goes
as follows. 
The Bogoliubov coefficients are still determined by taking  
the inverse Fourier transform of $\tilde \varphi_\lambda^{in}$, given 
by eq. (\ref{r12}) wherein one puts $A= 1/{\sqrt{4 \pi \lambda}}$, 
and $B=0$. Then, one sees that
for $\vert w \vert \gg 1/ \omega_c$,  the integral is dominated by
values of $p$ in the cis-Planckian domain $p \ll \omega_c$. 
Irrespectively of the nature of
corrections to the dispersion relation encoded in $g(p)$,
this locality implies that for
these `large' $\vert w \vert$, one recovers the un-modified
propagation on each side of the horizon which characterizes the out
modes. This in turn implies that
one also recovers  the usual Bogoluibov coefficients,
for more details see \cite{BMPS95}.

In order to discuss the third point one must choose the  deviation
from linearity in $g(p)$. For our  purpose, it is sufficient to
consider $g(p)$ of the form $g(p) = p ( 1 + \xi p^2/ \omega_c^2)$. 
For $\xi = 1$ one obtains sub-luminous propagation, for $\xi = -1$
super-luminous propagation and for  $\xi = i$ one gets dissipation
without significant dispersion. These results are easily reached by
constructing wave packets  and analysing the locus where their phase
is stationary. The main point is the following: when the phase of the
function varies faster (slower) than the unmodified phase given by
$\lambda \ln w$, one has sub-luminous (supra-luminous) propagation. 
In anticipation to what we shall get for metric fluctuations,
we say a few more words in the case of dissipation.
In this case, the first order deviation for $\lambda \gg1$ is of the form 
\ba
\tilde \varphi_\lambda^{in}(p) &=&  
{ 1 \over \sqrt{ 4 \pi \lambda}}
\left\{ { \exp \left( - i \lambda \int^p dp' { 1  \over 
p'(1 + i p'^2 /\omega^2_c) }\right) 
\over  p(1 + i p^2 /\omega^2_c) } \right\}
\nonumber\\
&\simeq& 
{  p^{-i\lambda -1 } \over \sqrt{ 4 \pi \lambda}}
e^{- \lambda p^2/2 \omega^2_c} \, ( 1 + O(p^2/\omega^2_c))
\label{disp}
\ea
This essentially corresponds to that will be obtained 
in the case of metric fluctuations.
The fact that dissipation (and not only dispersion)
should physically occur for black hole was discussed in \cite{BMPS95}
and the space time image of a packet built 
with waves of the type (\ref{disp})
 was schematically presented in Fig. 5, 
see also \cite{BS2s} for a super-luminous dispersion relation
which leads to an effective dissipation of the wave packet.  

There is one more point which should be discussed. It concerns the
physical relevance of considering propagation {\it backward} in time.
A few remarks may illustrate its relevance\footnote{However in the absence of a
manageable theory of quantum gravity,  we are still missing  an
explicite computation of backreaction effects which will settle this
question in unambiguous terms.}. First backward
propagation provides the simplest tool to investigate the nature of
the field  configurations which give rise to Hawking radiation. In
particular, it clearly establishes that Hawking quanta emerge from
trans-Planckian configurations when one uses, as Hawking originaly
did, the free Dalembertian for propagating the modes. Secondly, when
considering $S$-matrix elements in a quantum dynamical framework,
backward propagation always occurs since one also fixes the final
state of the field, see \cite{MP95, thooft96}. The simplest example is provided
the Feynman $in$-$out$ Green function.
This function can be easily related to $in$ modes characterized by an 
$out$-energy $\lambda$. Indeed, one has
\ba
\alpha^{-1}_\lambda  \left( \varphi_\lambda^{in}(w) \right)^* &=&
\int \!du \, e^{i \lambda u} \,{ <0,out \vert \varphi(u, v=\infty) 
\varphi(w, v=0 )\vert 0, in>
\over < 0,out \vert 0, in>}
\nonumber\\
&=& \int\! du \, e^{i \lambda u} \, G_{\ind{in-out}}(u, v=\infty ; w, v=0)
\label{5.31}
\ea

Because of its out-frequency content, the wave function
is global in the sense that it entails the propagation 
from ${\cal J}^+$ to the point where it is evaluated, 
here $(w=2 r -1, v=0)$.  In particular when computed in 
a fluctuating geometry, this function encodes the effects of
the metric fluctuations one encounters from ${\cal J}^+$ 
to the point where it is evaluated.

\subsection{Backward propagation of $out$-modes}

We now have all the tools to perform the comparison. To be as close
as possible to what we just presented, we consider the image on $v=0$
of the monochromatic wave of out-frequency $\lambda$.  That is, the
image on ${\cal{J}}^+$ is ${\mathbf \Phi}^+_{(\lambda)}(y)
=\varphi^{out}_\lambda(y=w)$  given in eq. (\ref{r15}). 

Then we have to face the problem mentioned after eq. (\ref{3.38b}):
$\varphi^{out}_\lambda(y)$ is not a smooth function but a distribution.
This is of course a manisfestation of the trans-Planckian problem. 
For instance, its derivative with respect to $y$ is ill defined on $y=0$.
 Therefore, one must regularize $\varphi^{out}_\lambda$
before applying the scattering operator ${\bf {\sf D}}$ on it. 

The simplest way to define the action of $\ll{\bf {\sf D}}\gg$ on our 
$out$-function is to work in the momentum conjugated to $w$.  Indeed,
the Fourier transform of $\varphi^{out}_\lambda$ is well defined
in the high $p$ regime, see eq. (\ref{r432}). Then using eq. (\ref{3.40}) we
simpy get 
\[
\ll \Phi^-_{(\lambda)}(p) \gg = \exp\left(- {p^2 \sigma^{2}_{\ind{eff}} \over 2}
\right) {\mathbf \Phi}^+_{(\lambda)}(p)
= \exp\left( - {p^2 \sigma_{\ind{eff}}^2 \over 2}
\right) {\Gamma(i \lambda + 1) \over {\sqrt{8 \pi^2 \lambda}} }
 (\epsilon + ip)^{-i \lambda - 1}
\]
\be
= \exp\left(- {p^2 \sigma^{2}_{\ind{eff}} \over 2}\right)
 { (-i) \Gamma(i \lambda + 1) \over {\sqrt{8 \pi^2 \lambda}} }
\left[ e^{\pi \lambda /2} \vartheta(p) p^{-i \lambda - 1}  -  
e^{-\pi \lambda /2} \vartheta(-p) (-p)^{-i\lambda -1} \right]
\, .\label{r10}
\ee
Since the
effect of the stochastic fluctuations is to multiply the wave function
by an even function in $p$,  the relative weight  encoding the Bogoliubov
coefficients is unaffected. This guarantees that the
vacuum state with respect to $p >0$ leads to the usual properties of
Hawking radiation, thereby proving point 2 above.
Moreover, eq. (\ref{r10}) confirms that the trans-Planckian 
 problem is tamed: The high frequency content, 
i.e. the near horizon behaviour, is suppressed by
a Gaussian factor, as for the {\it dissipative} case in the 
former Section, see eq. (\ref{disp}).
Finally, for $p^2 \sigma_{\ind{eff}} \ll 1$,
 one recovers  the usual expression of
out-modes in terms  of $p$ given in (\ref{r432}),
  thereby proving point 1 above.

To complete our analysis, we shall now determine the behaviour
of $\ll\!\Phi^-_{(\lambda)}\!\gg$ in
spacetime.  To this end, we inverse Fourier transform  separately the
two terms (positive and negative $p$) which appear in 
eq. (\ref{r10}).  Using eq. (1.11) in \cite{GO},
we obtain 
\be
\ll\Phi^-_{(\lambda)}(w)\gg =  { \sigma_{\ind{eff}}^{i \lambda} 
\over {\sqrt{4 \pi \lambda}} }
\, Z_{\lambda}(w/\sigma_{\ind{eff}})\, ,
\label{5.33}
\ee
where
\be
Z_{\lambda}(x)=
{e^{- x^2 / 4}
\over 2 \sinh(\pi \lambda)}
\left[ e^{\pi \lambda /2} 
D_{i \lambda } ( \epsilon -i x )
-  e^{ -\pi \lambda /2} D_{i \lambda } ( \epsilon + ix )
\right] \, .
 \label{r16}
\ee
Here $D_{\mu} (z)$ is the  parabolic cylinder function. 
This function is related to the Whittaker function $W_{\kappa,\nu}(x)$
(see, e.g. \cite{Buch:69}, p.39)
\be\n{wh}
D_{\mu} (z)=2^{\mu/2}\, \left({z^2\over 2}\right)^{-1/4}\,
W_{\mu/2+1/4,1/4}\left({z^2\over 2}\right)\, .
\ee
In (\ref{r16}) we have introduced $\epsilon$ positive and infinitesimal in order to
specify the phase of $\epsilon -ix$ for positive and
negative $x$.
\begin{figure}
\begin{tabular}{cc}
{\epsfig{file=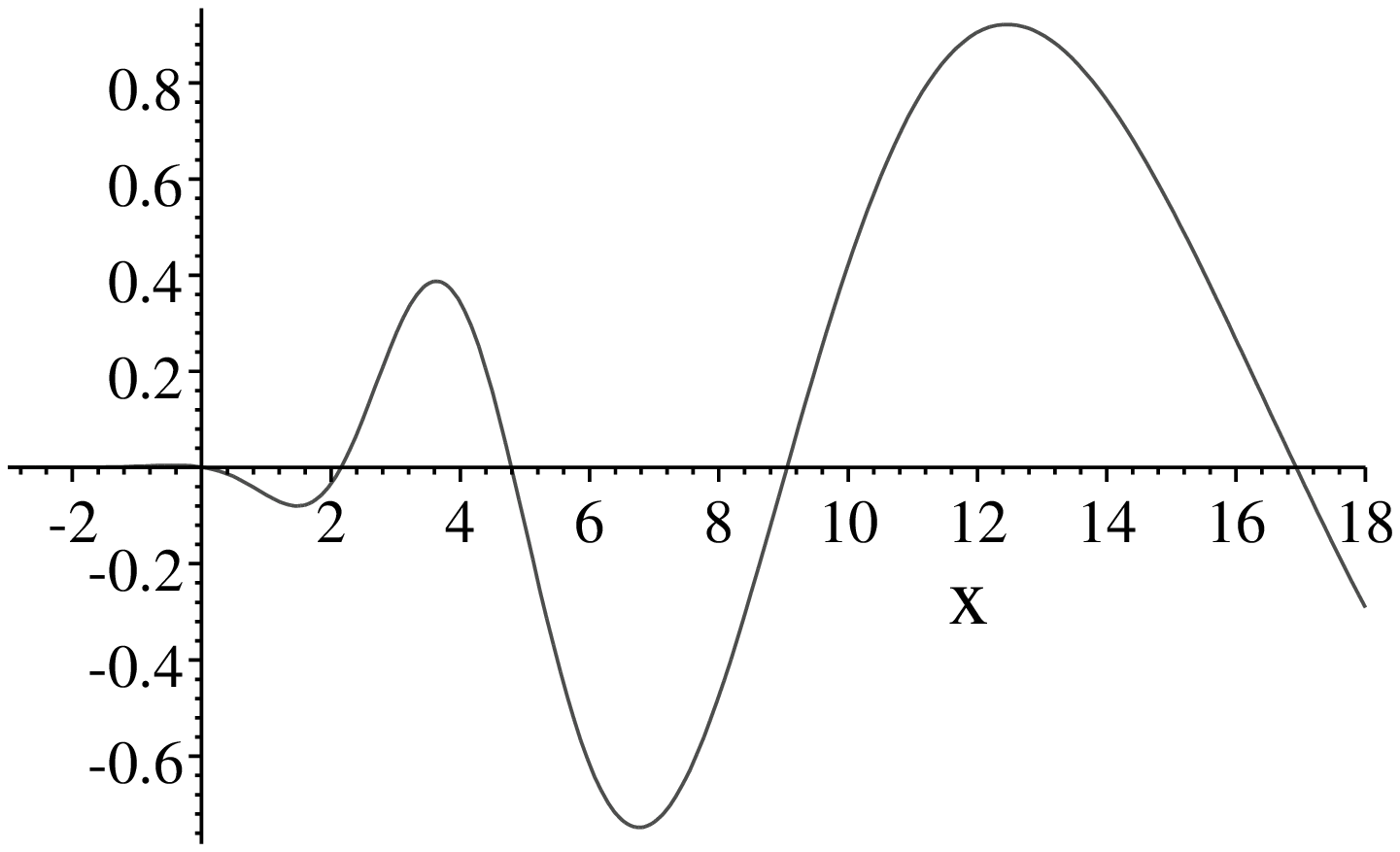, width=7.5cm}} &
{\epsfig{file=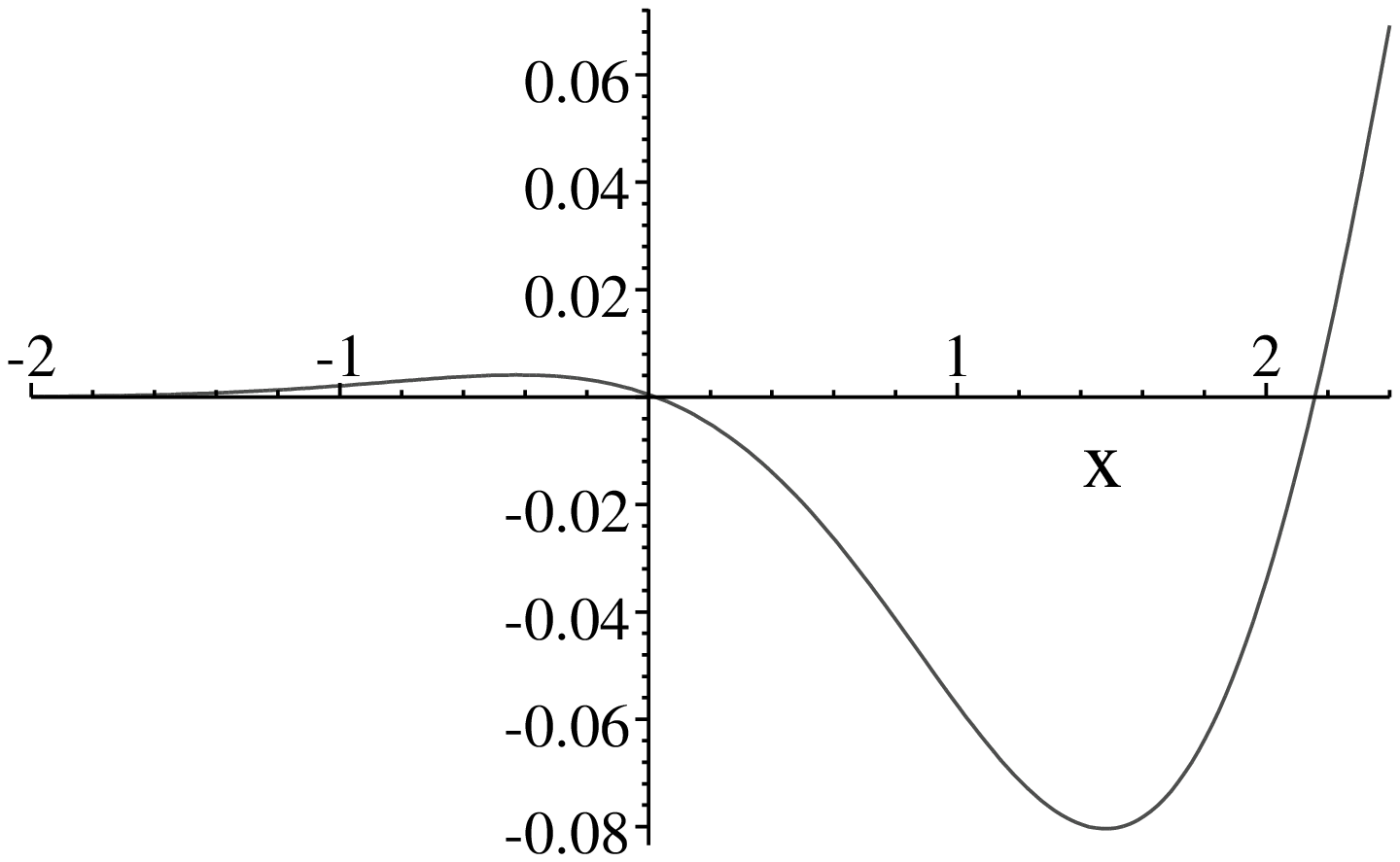, width=7.5cm}} \\
{$\bf (a)$} \hfill& {$\bf (b)$}\hfill\\
\end{tabular}
\caption[parcyl]
{Plot (a) gives the real part of $Z_{\lambda}(x)$ 
as a function of $x$ for $\lambda=5.0$.
Plot (b) gives an higher resolution of the same function 
in the interval $-2.0<x<2.0$.}
\label{parcyl}
\end{figure} 
Figure~\ref{parcyl} illustrates the behavior of
$Z_{\lambda}(x)$. We have plotted 
the real part of $Z_{\lambda}(x)$ using Maple
and relation (\ref{wh}). 

\begin{figure}[t]
\centerline{\epsfig{file=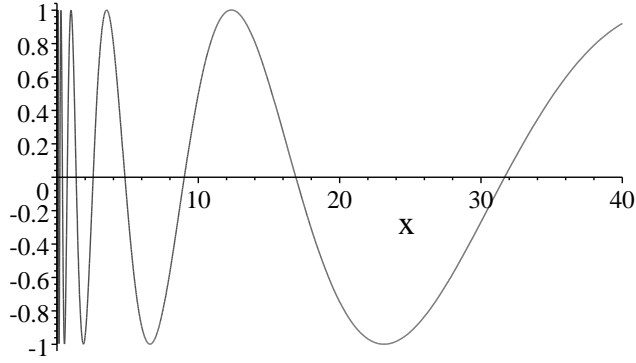, width=8.5cm}}\caption[unp]
{Plot of the real part of the unperturbed out-wave $x^{i\lambda}$
for $\lambda = 5$.}
\label{unp}
\end{figure}

The function 
$D_{i\lambda}(z)$ has the following asymptotic behavior
for $\vert z \vert \gg \vert \lambda \vert$, see \cite{Abram},
\be
D_{i\lambda }(z) \sim z^{i \lambda} e^{-z^2/4} \left\{
1 + { \lambda (\lambda + i) \over 2 z^2} + O(z^{-4})  \right\} \, , 
\hspace{0.5cm}\mbox{for}\,
-{\pi\over 2}<\mbox{arg}\, z <{\pi\over 2}\, 
\hspace{0.5cm}
.
\ee
Using this relation, one verifies that for large 
negative values of $x=w/\sigma_{\ind{eff}}$, 
$Z_{\lambda}$ vanishes. 
Instead, for large positive $z$ behaves as
\be\n{as}
Z_{\lambda}(x)\sim x^{i\lambda} \left\{
1 - { \lambda (\lambda + i) \over 2 x^2} + O(x^{-4})  \right\}
\, , 
\ee
The limit $x=w/\sigma_{\ind{eff}} \rightarrow \infty$ corresponds to
the far from horizon region or to
$\sigma_{\ind{eff}}\rightarrow 0$. The last case corresponds to no
metric fluctuations. In this regime (\ref{as}) reproduces the
unperturbed out wave function given in  eq. (\ref{r15}). The plot of
the unperturbed out-wave function $x^{i\lambda}$ is shown in Figure~\ref{unp}. 
By comparing Figures~\ref{parcyl} and \ref{unp} one clearly sees that
metric fluctuations strongly affect the back
scattered wave for values of $w$ close to smeared horizon. 
Indeed, close to the horizon, i.e. for $x^2 \ll \lambda$, 
 the function $Z_{\lambda}$ behaves as, see \cite{Abram},
\be\n{zero}
Z_{\lambda}(x)=
{2^{i\lambda/2}\,\over 2\sqrt{\pi}}\,  \Gamma\left({1+i\lambda \over 2}  \right)
\left\{ 1 + \coth( \pi \lambda /2) e^{i \pi /4} x \sqrt{\lambda - i/2} 
+ O( x^2 \lambda)
\right\} 
\ee
The most important feature is the disappearance of the infinite
trans-Planckian reservoir of fluctuations of equal amplitude which 
characterizes the unperturbed out-wave $w^{i \lambda}$. 
This implies that backward propagation {stops} for any localized wave packet. 
In order words, the amplitude of the wave packet is dissipated 
in its backward motion once the wave packet enters the near
horizon region $x< \lambda$.

To show this, let us consider a rapidly modulated Gaussian wave
packet which on ${\cal J}^{+}$ has the image 
\be\n{gwp}
\Phi^+_{\bar \lambda}(u)= e^{-i \bar \lambda(u - u_0)} 
e^{-(u-u_0)^2 \over 2b^2}\, .
\ee
This image is localized around $u=u_0$ and has width $b$. In the limit
$b \to \infty$ one obtains a monochromatic out-plane wave 
of frequency $\bar \lambda$. Its Fourier decomposition is given by 
\be
\Phi^+_{\bar \lambda}(u)= {b\over \sqrt{2\pi}}\,
\int_{-\infty}^{\infty}\, d\lambda\, e^{-i(u-u_0) \lambda}\,
e^{-b^2 (\lambda - \bar \lambda)^2/2}\, .
\ee
We first consider its backward propagation in the usual non-fluctuating
geometry. In this case, the plane wave $e^{-i\lambda u}$ takes the form
$w^{i\lambda}$ on $v=0$.  Thus the image on $v=0$ of the wave-packet
(\ref{gwp}) is
\be\n{nonper}
\Phi^-_{\bar \lambda}(w)= 
e^{i \bar \lambda(\ln w + u_0)}
e^{-(\ln w +u_0)^2/(2b^2)}\, .
\ee
For late time $u_0$, it hugs the horizon $w=0$ since 
it is centered around $w= e^{-u_0}$. 
Moreover it is highly blue-shifted since
its mean frequency is equal to $\bar \lambda e^{u_0}$ 
for an infalling observer. Finally, no matter how large $u_0$ is,
its maximum amplitude is still 1. There is no dissipation. 
This is the trans-Planckian problem. 

Let us now analyze how the fluctuating geometry changes this picture. 
In the presence of fluctuations, the image on $v=0$ of the plane wave
$e^{-i\lambda u}$ is $\sigma_{\ind{eff}}^{i\lambda}\,
Z_{\lambda}(w/\sigma_{\ind{eff}})$ instead of $w^{i\lambda}$. For
simplicity, we first consider a wave packet with $\bar \lambda \gg 1$
and  $b \gg 1$. The first condition means that it is rapidly
oscillating (in the units of the surface gravity)  whereas the second
means that its frequency content is well peaked around the mean
frequency $\bar \lambda$. Using the near horizon behavior of
$Z_{\lambda}$,  eq. (\ref{zero}), the scattered wave packet is
\ba
\ll\!\Phi^-_{\bar \lambda}(w)\!\gg &=& {b\over \sqrt{2\pi}}\,
\int_{-\infty}^{\infty}\, d\lambda\, e^{i u_0\lambda}\,
e^{-b^2(\lambda - \bar \lambda)^2/2} \sigma_{\ind{eff}}^{i\lambda}\, 
Z_{\lambda}(w/\sigma_{\ind{eff}})  \,
\label{ZZ}\\
&\simeq&
{b\over \sqrt{8 \pi}}\,
\int_{-\infty}^{\infty}\, d\lambda\, e^{i \lambda(u_0 + 
\ln \sigma_{\ind{eff}} + \ln(\lambda)/2)}\,
e^{-b^2(\lambda - \bar \lambda)^2/2} \left\{ 1 +  
{ w \over \sigma_{\ind{eff}}}
e^{i \pi /4} \sqrt{\lambda} 
\right\}\, .
\nonumber
\ea
In the second line we have dropped all irrelevant phases and we have
used the asymptotic behavior of the $\Gamma$ function and the $\coth$,
in anticipation to the fact that the main contribution of the integral
will arise from values of $\lambda$ centered around $\bar \lambda$.

By performing the integral by a saddle point approximation
(this is perfectly valid when $b^2 \gg 1$), one obtains the following behavior
\be
\ll\!\Phi^-_{\bar \lambda}(w)\!\gg  \simeq 
\exp \left(- { (u_0 + \ln \sigma_{\ind{eff}} + {1 \over 2} \ln \bar \lambda )^2 
\over 2 b^2 }\right)
\left\{ 1 +  {w \over \sigma_{\ind{eff}} }
e^{i \pi /4} \sqrt{\bar \lambda} 
\right\}\, .
\ee
Hence once $u_0 > - \ln \sigma_{\ind{eff}}$, the image on $v=0$
vanishes like a Gaussian in $u_0$.

This strong dissipation is only valid for tight wave packets in $\lambda$,
i.e. for $b \gg 1$.  Instead, in the opposite regime $b \ll 1$,
for tight wave packets in position space, the dissipation is milder. 
Indeed, in the limit $b \to 0$, the Gaussian factor in eq. (\ref{ZZ}) 
can be ignored in the large $u_0$ limit. Then, the main
contribution comes from the first pole of the  $\Gamma$ function in the
positive imaginary $\lambda$ axis. In this regime the decrease of the
wave packet is given by $e^{- (u_0 + \ln \sigma_{\ind{eff}})}$. 

In brief, as long as the mean position in  $w$ on $v=0$ of the wave
packet is  larger than $\lambda \sigma_{\ind{eff}}$, its image is
unaffected by the  metric fluctuations since $Z_\lambda$ still behaves
as $w^{i \lambda}$. Instead, once it enters the near horizon region,
its amplitude rapidly decreases.

\section{Conclusion}

It is perhaps appropriate to list what we learn from our analysis.

\begin{enumerate}
\item When the relative width of the smeared horizon is small enough, 
i.e. when $\delta r_{EH}/r_{EH} \simeq  \sigma_{\ind{eff}} \ll 1$,
metric fluctuations in the near horizon geometry affect the asymptotic
properties of Hawking radiation only slightly, in the second order of
$\sigma_{\ind{eff}}$, see (\ref{4.18}).
\item The reason for this stability can be seen from the 
short distance behaviour of the {\it in}-Green function, see (\ref{4.7}).
Indeed, for $|\Delta y |\ll 1$, one recuperates the usual 
Hadamard behavior which guarantees point 1.
\item Backward propagated wave packets representing Hawking quanta 
of energy $\lambda$ are dissipated when their Doppler shift frequency
$i\partial_r$ reaches $\sigma_{\ind{eff}}^{-1}$, i.e. when their separation
in $r$ from the event horizon approaches $\lambda \sigma_{\ind{eff}}$,
see (\ref{ZZ}).
\end{enumerate}
The attentive reader will notice that points 2 and 3
are rather difficult to conciliate. Indeed if the behavior of the
 in-Green function is not modified why for is point 3 relevant?
In other terms what is the high energy behavior of the theory?
Unmodified as suggested by point 2 or dramatically modified
as indicated by point 3?

To answer these questions one should first  reconsider the domain of
validity of the scheme we used. Our scheme is based on eq. (\ref{3.2})
which represents the free propagation of a test-field in a fluctuating
geometry. This is an approximative description which completely
neglects the back-reaction effects induced by the field $\varphi$
itself. In other words we are working in the regime when the  metric
fluctuations are not significantly affected by the  energy density
carried by $\varphi$. Therefore, our main hypothesis is that there is
an {\it intermediate} regime in which  the metric fluctuations induced
by all the other degrees of freedom cannot be neglected whereas the
back-reaction effects due to $\varphi$  can be neglected. For
frequencies below this regime, we have proven that  the metric
fluctuations play no role, see eq. (\ref{3.39}). For higher
frequencies, we cannot say much.  Nevertheless it is most probable that
the stochastic description we used fails. Therefore we can trust the
behavior of the in-Green function only for separations within the
intermediate domain.

Having clarified this point, we can now explain why  the high frequency
behavior of the in-Green function differs so much from that of backward
scattered waves. The reason is the following.  Being a function of the
difference in $V_\phi$, the in-Green function is hardly sensitive to
the metric fluctuations in the coincidence point limit, see eq.
(\ref{4.7}). On the contrary, as emphasized after eq. (\ref{5.31}),
backscattered wave functions defined on ${\cal J}^+$ are sensitive to
the metric fluctuations they have encountered when evaluated near the
horizon. Moreover, since their frequency is blue shifted, they are
inevitably strongly affected by the metric fluctuations. 

If this explains that the behaviour of the in-Green
is perfectly compatible with that of backscattered waves, 
it does not tell us what happens to these waves and their energy 
density when their amplitudes diminish. To describe 
the fate of their energy density one must consider 
the dynamics of the degrees of freedom which engender the metric
fluctuations. Indeed, one must go beyond the stochastic treatment
of these fluctuations in order to be able 
to describe the trans-Planckian momentum recoils which shall
{\it inevitably} be induced by Hawking photons when they are 
traced backward near the event horizon, see 
\cite{thooft96, feyn} for preliminary attemps to describe this physics.

In brief, the main outcome of the paper is to have provided physical
foundations in terms of metric fluctuations to the concept of effective
propagation of light near a black hole horizon.

This allows to address in a rational scheme the question of the domain
of validity of this effective propagation. It also provides an
explanation for the vexing question of the apparent violation of 
local Lorentz invariance\cite{Jac91, Jac93, BMPS95}. 
The neatest way to characterize this violation is
to focus on the near horizon behavior of a monochromatic mode
$\phi_\lambda^{out}$. In the absence of modification of the dispersion
relation, this mode behaves as $w^{i \lambda}$ where $w=2r - 1$. Hence
there is no length which allows one to distinguish low from high
momenta. This absence is a consequence of the local Lorentz invariance
of theories based on the usual Dalembertian. 
On the contrary, when dealing with a modified dispersion
relation, one breaks this invariance since the new dynamical 
equation is written in a preferred frame.  For acoustic black
holes this makes good sense since both the frame and the critical lenght, 
which characterizes what ``high'' frequency means,
are given by the constituents of the fluid. On the contrary, it is
rather unclear to see the origin of such a preferred frame 
for a gravitational black hole.  One of the main virtues of the
present work  is to provides a simple answer to this puzzle. 
Indeed the ensemble of metric fluctuations unambiguously determines, 
$\sigma$, the constant spread in $r$ (measured along $v=0$) of the
distribution  of the backward propagated rays representing the event
horizon. Because of the hypothesis of stationarity metric fluctuations,
the modified equation governing light propagation has a simple and
stationary expression in the $v, r$ coordinate system. In particular,
the cut-off lenght $\sigma$ appears only through  powers of $\sigma
\partial_r \vert_v$.  In this case what might be interpreted  as the
origin of a ``violation of Lorentz invariance'' originates from the
ensemble of stationary metric fluctuations.

\bigskip

\vspace{12pt}
{\bf Acknowledgments}:\ \  This work was  partly supported  by the
Centre National de la Recherche Scientifique (France) and was done in
the framework of a NATO collaboration (CRG.972079). One of the authors (V.F.) is
grateful to the Natural Sciences and Engineering Research Council of
Canada and to the Killam Trust for their financial support.  R.P.
wishes to thank Robert Brout for many discussions in the
search for an effective description of the trans-Planckian soup.

\bigskip

\appendix

\section{The leading non-linear corrections}
\setcounter{equation}0

In this Appendix we shall demonstrate the validity of using the linearized
and simplified expression, eq. (\ref{2.20}), of $V_\phi(u)$ for 
obtaining the dominant non-linear effects in $w_0$
on the propagation of the field described by eq. (\ref{3.9}).

To simplify the analysis, 
we consider the Fourier transform with respect to 
$w$ of the image of the pulse defined on ${\cal J}^-$
evaluated in a given realization:
\be
\Delta_\phi ^+(u \vert p) = {1 \over \sqrt{2\pi}} \int dw\, e^{ipw} \, 
\Delta_\phi^+(u \vert w) \, .
\ee

Using eq. (\ref{3.7}), the mean image is given by
\be \n{A.2}
\bar \Delta ^+(u \vert p) = {1 \over (2\pi)^{3/2}}\int_0^{ 2 \pi}\! d\phi \, 
e^{ip V_\phi(u)} \, ,
\ee
the integral over $w$ being trivial.

The analysis of the null ray propagation in a fluctuating geometry
shows \cite{BaFrPa:99} that the complete expression for $V_\phi(u)$ is of the form
\be\n{A.3}
V_\phi(u)= -1 - e^{-u} + F_0 +F_1 e^{-u} + O(e^{-2u})\, ,
\ee
where the correction terms $F_0$ and $F_1$ are series 
in $w_0$. The corrections terms multiplied by $e^{-u}$ or higher powers
of it are not important in the late time regime since 
they become arbitrarily smaller than those of $F_0$. Hence we can drop
$F_1$. The remaining correction term $F_0$ fixes 
the precise value of the image on ${\cal J}^-$
of the exact horizon in every realization of the fluctuating geometry.
It also determines the late times corrections to $\bar \Delta ^+(u \vert p)$.
It has the following expansion in powers of $w_0$ 
\be\n{A.4}
F_0= w_0 \sin (\phi+ \phi_0) +w_0^2 f_1(\omega)\sin(2\phi
+\phi_1)+O(w_0^3)\, .
\ee
Let us substitute (\ref{A.4}) into (\ref{A.2}) and show that the
leading non-linear corrections to $\bar \Delta ^+$ all come 
from the linear term in $w_0$.

In the low $p$ regime, i.e. for $p \ll 1/w_0$, all correction terms are 
negligeable since $w_0 \ll 1$ in our dimensionless units.
Upon reaching the high $p$ regime, for $p \simeq  1/w_0$,
all terms of the form $p w_0^n$ for $n\ge 2$ are still very small. 
Using this fact we can rewrite (\ref{A.2}) as
\[
\bar \Delta ^+(u \vert p) = {e^{-ip ( 1 + e^{-u})}
 \over (2\pi)^{3/2}} \times
\int_0^{ 2 \pi}\! d\phi \,  e^{ipw_0 \sin\phi}\left[ 1+ 
i p w_0^2 f_1(\omega)\sin(2\phi
+{\phi}_1) +\ldots \right]
\]
\be
= {e^{-ip ( 1 + e^{-u})}  \over (2\pi)^{3/2}} \times \left[
G_0(pw_0)+ p w_0^2 G_1(pw_0) +p w_0^3 G_2(pw_0)+\ldots 
\right] \, .
\ee
The form of the integrals implies that  functions
$G_n(p w_0)$ are regular. For $pw_0 =0$ they are
all finite, while for $pw_0\to\infty$, $G_n(pw_0)$ are of the same order
of magnitude. Therefore the corrections to the leading term $G_0$ 
remain small as long as $pw_0^2\ll 1$. 
If $w_0$ is of the order of the Planck length,
 the regime of validity of the approximation, 
$p\ll w_0^{-2}$, goes far beyond the Planckian scale.
Having established this result, we can now make more precise 
the neglection of the $F_1$ term in eq. (\ref{A.3}) in the late
time regime. It is sufficient to have $e^{-u} < w_0$,
since the linear term in $w_0$ in $F_1$
behaves in this regime like the quadratic term of $F_0$.

In other words, at late times and for all the regimes starting from small $p$
till momenta much higher than the Planck one,
$G_0(pw_0)$ ($=2 \pi J_0 ( p w_0)$, see (\ref{3.28}))
gives the leading non-linear
corrections to $\bar \Delta ^+(u \vert p)$. This proves the claim
since $G_0(pw_0)$ entirely comes from the linear term in $F_0$.

\section{Backward Scattering}
\setcounter{equation}0

In this Appendix we discuss the properties of the backscattered
wave of out-frequency $\lambda$ 
when the average is performed over the phase only. 

Its image on ${\cal J}^-$ or on $v=0$,  $\bar \Phi^-_{(\lambda)}$,
 can be calculated by making a Fourier transform as it was done for
$\ll \! \Phi^-_{(\lambda)}(p)\!\gg$ in Section~5. Here, we
shall offer an alternative way to compute this image. 
We start by regularizing the out wave $\varphi^{out}_\lambda(y)$,
see eq. (\ref{r15}). One can view
this distribution as the limit $\epsilon \to 0$ on the real $w=y$ axis 
of the difference of two analytic functions. Explicitly
one has 
\be 
\sqrt{4 \pi \lambda} \varphi^{out}_\lambda(w) = \vartheta(w) w^{i \lambda} 
=
{ 1 \over 2 \sinh( \pi \lambda)}
\left( 
e^{\pi \lambda /2} (\epsilon - iw)^{i \lambda} 
- e^{-\pi \lambda /2} (\epsilon + iw)^{i \lambda}
\right) 
\ee
Notice that the out-mode has been expressed by a Bogoliubov  relation
as  the difference of two in-modes which are perfectly well defined on
the horizon.  We can then safely apply $\bar{\bf {\sf D}}$ to each of
them.  We shall make use of the relation
\be
P_{\nu}^{-\mu}\left( {z\over \sqrt{\rho^2+z^2}}\right)={\Gamma(\nu-\mu+1)\over
\Gamma(\nu+1) (\rho^2+z^2)^{\nu/2}} J_{\mu} \left(\rho {\partial \over
\partial z}\right)\, z^{\nu}\, ,
\label{Hist}\ee
where $P_{\nu}^{-\mu}$ denotes  the generalized Legendre function. Eq.
(\ref{Hist}) was derived by Filon in 1903 \cite{Filo:03} and cited  in
the book by Watson \cite{Wats:66}, p.51,  who proposed to the reader to
prove this formula as an exercise!
 Using this equation, we obtain 
\ba
&&\bar \Phi^-_{(\lambda)}(w) = J_0( i w_0 \partial_w ) \varphi^{out}_\lambda (w)
\nonumber\\
&&\quad \quad \quad = { 1 \over {\sqrt{4 \pi
\lambda}} }
{ 1 \over 2 \sinh(\pi \lambda)}
\left\{ e^{\pi \lambda /2} 
\left[(\epsilon -i w )^2 + w_0^2 \right]^{i \lambda /2}
P_{i \lambda } \left( {\epsilon -iw  \over 
\sqrt{(\epsilon -iw  )^2 + w_0^2 }}\right)
\right.
\nonumber\\
&&
\left.
\quad  \quad  \quad  \quad  \quad  \quad  \quad  
 \quad  \quad  \quad  \quad  \quad  \quad  
-  e^{ -\pi \lambda /2}\left\{ -iw \to +iw \right\}\right\}\, ,
 \label{r16b}
\ea
where $P_{\mu} (z)$ is the Legendre function. We have
introduced $\epsilon$ positive and
infinitesimal in order to specify
 the phase of $\sqrt{(\epsilon -iw  )^2 +
w_0^2 }$ in the three sectors: $\vert w \vert < w_0 $ and $\vert w
\vert > w_0 $.

Before presenting the detailed behavior of $ \bar \Phi^-_{(\lambda)}$
a few simple deductions can be made. First, in the limit $w_0 \to 0$,
$ \bar \Phi^-_{(\lambda)}$ reduces to the usual out wave function
given in  eq. (\ref{r15}) since $P_{i \lambda }(1)= 1$. 
Secondly,  one finds, as one
should do, that $\bar \Phi^-_{(\lambda)}(w) $ vanishes for $w < w_0$
since ${\mathbf \Phi}^+_{(\lambda)}(y)$ vanished for negative $y$. 
Thirdly,  the spread of the effective horizon $w = \pm w_0$ clearly
appears in $ \bar \Phi^-_{(\lambda)}$.

\begin{figure}
\begin{tabular}{cc}
{\epsfig{file=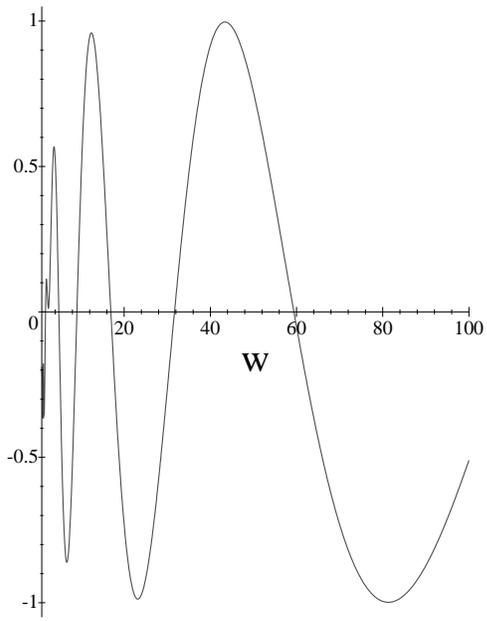, width=6.5cm}} &
{\epsfig{file=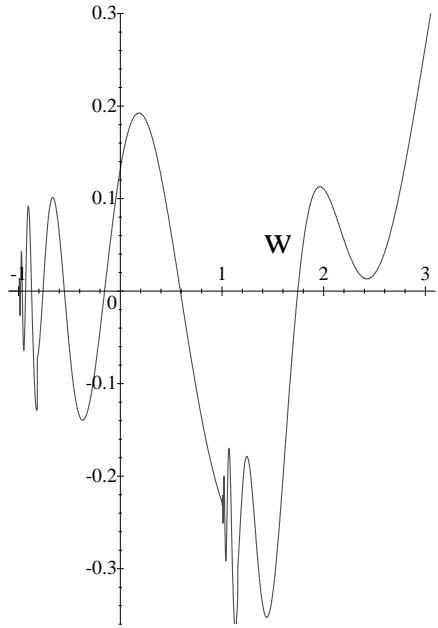, width=6.5cm}} \\
{$\bf (a)$} \hfill& {$\bf (b)$}\hfill\\
{\epsfig{file=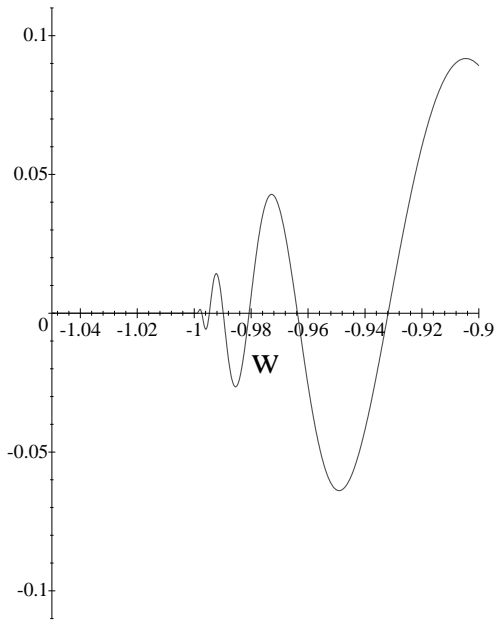, width=6.5cm}} &
{\epsfig{file=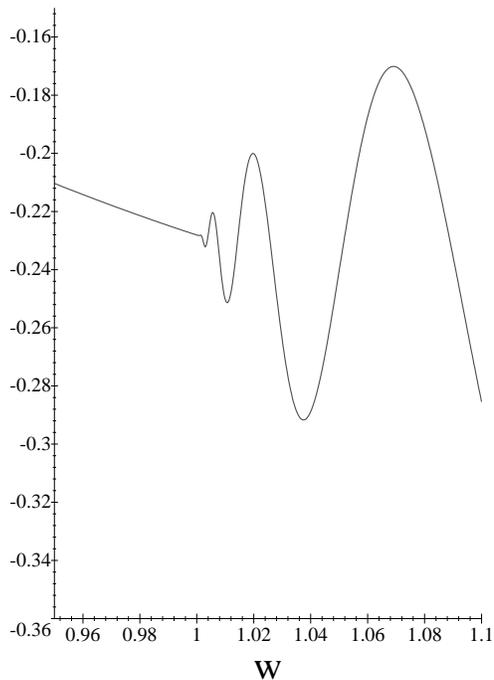, width=6.5cm}}\vspace{0.3cm}\\
{$\bf (c)$} \hfill& {$\bf (d)$}\hfill
\end{tabular}
\caption[Phi]
{These figures illustrate the salient properties of $\bar{\Phi}^-_{(\lambda)}(w)$. 
We have plotted the real part of  $\sqrt{4\pi\lambda}\bar{\Phi}^-_{(\lambda)}$ 
as a function of $w/w_0$ for $\lambda=5.0$ for 4 different sectors of $w$.}
\label{Phi}
\end{figure} 

Figure~\ref{Phi} illustrates  the behavior of $ \bar
\Phi^-_{(\lambda)}$ for $\lambda =5$. 
This behavior is typical for values $\lambda$ greater than $1$. 
For smaller frequencies, the amplitude of the oscillations
between $\pm w_0$ is much smaller than $1$. We have presented
only the real part of $ \bar \Phi^-_{(\lambda)}$ since
the imaginary part behaves similarly. 

The main features of $ \bar \Phi^-_{(\lambda)}$  directly follow from 
its properties near the singular points $w=\infty$ and $w=\pm
w_0$. The first manifestation of the metric fluctuations shows up
around $w = 20 w_0$ where the amplitude of the oscillations start to
decrease, see (fig. 5a). 
This modulation of the  amplitude arises from the
correction term of the asymptotic behavior of $ \bar
\Phi^-_{(\lambda)}$. Using eq. 8.1.2 in \cite{Abram}, one has
\be
\bar \Phi^-_{(\lambda)}(w) = { 1 \over {\sqrt{4 \pi
\lambda}} } w^{- i \lambda} \left\{
1 - { \lambda (\lambda + i) \over 4 } \left(
{ w_0 \over w } \right)^{2} + O \left(
{ w_0 \over w } \right)^{4} 
\right\}
\label{nn1}
\ee

The second effect is located for values of 
$w$ slightly bigger than $w_0$, see (fig. 5b). One has a rapidly oscillating
function with a decreasing amplitude. The origin of this 
structure can be understood from eq. (\ref{3.9}):
the logarithmicly divergent behavior of $\varphi_\lambda^{out}$ at 
$w=0$ has been shifted with some coherence to $w=w_0$
because of the stationarity of the shift with respect to $\phi$ 
at its maximal value.
This behavior can be seen from the analytical property
of the function. Using eq. 8.1.5 in \cite{Abram},
one finds that $ \bar \Phi^-_{(\lambda)}$ is a sum of two
hypergeometric functions. The first one controls the overall
shape of the function near $w=w_0$ whereas the second one
provides the rapidly oscillating behaviour. In the limit $w \to w_0$
one has
\be
\bar \Phi^-_{(\lambda)}(w) = { 1 \over {\sqrt{4 \pi
\lambda}} }  w^{- i \lambda} \left\{
A_1 + A_2 \left( 1 - { w^2 \over w_0^2  } \right)^{1/2 + i \lambda} 
+ O \left( 1 - { w^2 \over w_0^2 } \right)
\right\}
\label{nn2}
\ee
where $A_1 = 2^{i \lambda} \pi^{-1/2} \Gamma(i \lambda + 1/2 )  
\Gamma^{-1}(i \lambda + 1)$ and $A_2 =  2^{-i \lambda - 1 }
\Gamma(- i \lambda - 1/2 ) \Gamma^{-1}(i \lambda )$.
Notice that one recovers arbitrarily high frequencies in this
behavior.  Therefore one can fear that one has simply shifted the 
trans-Planckian problem from $w=0$ to $w=w_0$. This is not the case
for two reasons. First the amplitude of these oscillations decreases
as $\sqrt{ w^2 - w_0^2}$ for $w \to w_0$. Therefore the amplitude to
find high frequencies decreases. Secondly, when one averages over
the amplitude $w_0$, this structure is erased, as shown in Figure 4.
Therefore, it is an artifact of our simplified averaging procedure.

The third salient fact is that on the other side 
of $w_0$ no rapid oscillations are found, see (figs. 5b-c). 
There are nevertheless oscillations but, as indicated in the next equation
(obtained again from  eq. 8.1.5 in \cite{Abram})
 they are suppressed by $\sinh (\pi \lambda)$ with $\lambda =5$.
\be
\bar \Phi^-_{(\lambda)}(w) = { 1 \over {\sqrt{4 \pi
\lambda}} }  w^{- i \lambda} \left\{
A_1 + i { A_2 \over \sinh(\pi \lambda)} 
\left( { w^2 \over w_0^2  } -1  \right)^{1/2 + i \lambda} 
+ O \left( { w^2 \over w_0^2 } -1 \right)
\right\}
\label{nn3}
\ee

The fourth point is that $\bar \Phi^-_{(\lambda)}$
is regular around $w=0$, see (fig. 5d). Analytically one has
\be
\bar \Phi^-_{(\lambda)}(w) = { 1 \over {\sqrt{4 \pi
\lambda}} }  w_0^{- i \lambda}
\left\{ \tilde A_1 + \tilde A_2 { w \over w_0}
+ O \left( { w^2 \over w_0^2 } \right)
\right\}
\label{nn4}
\ee
where $ \tilde A_1 = \pi^{-1/2} \Gamma ({1 \over 2} + {{i\lambda} \over 2})
\Gamma^{-1}(1 + {{i\lambda} \over 2})$ and 
$ \tilde A_2 = -{{i\lambda} \over 2}  \pi^{-1/2} \Gamma ({{i\lambda} \over 2})
\Gamma^{-1}({1 \over 2} + {i\lambda \over 2})$. 
Therefore, the trans-Planckian reservoir 
of oscillations which was present in $\varphi_\lambda^{out}$
has been eliminated. This is the main physical result. 

The fifth point concerns the structure when $w$ approaches
$-w_0$ from above.
Analytically, one has
\be
\bar \Phi^-_{(\lambda)}(w) = {  1 \over {\sqrt{4 \pi
\lambda}} }  (-w)^{- i \lambda} \left\{ - i { A_2 \over \tanh(\pi \lambda)} 
\left( { w^2 \over w_0^2  } -1  \right)^{1/2 + i \lambda} 
+ O \left( { w^2 \over w_0^2 } -1 \right)
\right\}
\label{nn5}
\ee
As in eq. (\ref{nn2}), the amplitude of the fluctuations deceases like
a square root when approaching the boundary $w=-w_0$. 
As already mentioned, for smaller values of $w$ one finds identically zero. 
One can apply to this fifth point the remarks
made in the second point.

In brief, the main physical result is the fourth point. Moreover,
in contradistinction with the second, third and fifth points,
 it is stable if one considers an ensemble of amplitudes $w_0$.
The physical consequence of the disparition of the reservoir 
of oscillations for $w< w_0$ is that the backward propagation 
of any wave packet will stop around $w_0$. 
See Section 5.2 for more details.


\begin{thebibliography}{9}

\bibitem{Hawk:75} S.W. Hawking, Comm. Math.
Phys. {\bf 43} (1975) 199.

\bibitem{Hu:99} B. L. Hu. Preprint   gr-qc/9902064.

\bibitem{CaHu:98} A. Campos and B. L. Hu. Preprint gr-qc/9812034.

\bibitem{MaVe:98a} R. Mart\'{i}n and E. Verdaguer. Preprint
gr-qc/9811070.

\bibitem{MaVe:98b} R. Mart\'{i}n and E. Verdaguer. Preprint
gr-qc/9812063.

\bibitem{York:83} J. W. York Jr., Phys. Rev. {\bf D28} (1983) 2929.

\bibitem{BaFrPa:99} C. Barrab\`es, V. P. Frolov and R. Parentani,
Phys. Rev. {\bf D59} 124010 (1999). 

\bibitem{Unruh81} W. Unruh, Phys. Rev. Lett {\bf 46} (1981) 1351

\bibitem{Jac91} T. Jacobson, Phys. Rev. D{\bf 44} (1991) 1731

\bibitem{Jac93}T. Jacobson, Phys. Rev. {\bf 48} (1993) 728

\bibitem{tHooft} C. R. Stephens, G. 't Hooft and B. F. Whiting,
Class. Quant. Grav. {\bf 11} (1994) 621.

\bibitem{VV} Y. Kiem, H. Verlinde, E. Verlinde, Phys. Rev. D {\bf 52}
(1995) 7053

\bibitem{MP95} S. Massar and R. Parentani, Phys. Rev. D {\bf 54} (1996) 7444

\bibitem{thooft96} G. 't Hooft, Int. J. Mod. Phys. A11 (1996) 4623

\bibitem{Unruh95} W. Unruh,  Phys. Rev. D {\bf 52} (1995) 2827

\bibitem{BMPS95} R. Brout {\it et al}, Phys. Rev. D {\bf 52} (1995) 4559

\bibitem{Jac96} T. Jacobson, Phys. Rev. D {\bf 53}  (1996) 7082

\bibitem{CEMNP} A. Casher {\it et al}, Nucl. Phys. B484, (1997) 419 

\bibitem{HuSh:98} B. L. Hu and K. Shiokawa, Phys. Rev. D {\bf 57} (1998)
3474.

\bibitem{WuFo:99} C.H. Wu and L.H. Ford, Phys.Rev.D {\bf 60} 
(1999) 104013.

\bibitem{FH}  R. Feynman and A. Hibbs, {\em Quantum Mechanics and
Path Integrals}, McGraw-Hill, New York, 1965.


\bibitem{VanK} N. G. van Kampen,  {\em   Stochastic processes in Physics
and Chemistry}, North Holland, Elsevier (1981)

\bibitem{FrNo:98} V. Frolov and I. Novikov. {\it Black Hole Physics:
Basic Concepts and New Developments,} Kluwer Academic Publ., 1998.

\bibitem{thoooft} G. 't Hooft Nucl. Phys. B {\bf 256} (1985) 727

\bibitem{Filo:03} L.N.G. Filon, Phil.Mag. (6) VI (1903), 193.

\bibitem{Wats:66} G.N. Watson, {\em A Treatise on the Theory of Bessel
Functions}, Cambridge Univ. Press, 1966.


\bibitem{BiDa:82} N.D. Birrel and P.C.W. Davies, 
{\it Quantum Fields in Curved Space},
Cambridge University Press, 1982.

\bibitem{HAA} K Fredenhagen and R. Haag, Commun. Math. Phys.
{\bf 127} (1990) 273


\bibitem{MP952} S. Massar and R. Parentani, Phys. Rev. D {\bf 54} (1997) 7426


\bibitem{GO} R. Brout {\it et al}, Phys. Rep. {\bf 260} (1995) 329

\bibitem{Visser} M. Visser, Class.Quant.Grav. 15 (1998) 1767

\bibitem{Reznik} B. Reznik, gr-qc/9703076

\bibitem{Jac99} T. Jacobson and D. Mattingly, hep-th/9908099, 

\bibitem{BS2s} R. Brout {\it et al}, Phys. Rev. D {\bf 59} (1999) 044005
 

\bibitem{feyn} R. Parentani, gr-qc/9904024, to appear in Phys. Rev. D61


\bibitem{GrRy:94} I.S. Gradshtein and I.M. Ryzhik. {\it Table of
Integrals, Series, and Products}, Academic Press, New York, 1994. 

\bibitem{Abram} M. Abramowitz and I. Stegun, {\it Handbook of
Mathematical Functions}, Dover, New York (1970) 

\bibitem{Buch:69} H. Buchholz. {\it The Confluent Hypergeometric
Function}, Springer-Verlag, New York Inc., 1969.













\end{thebibliography}
\end{document}